\documentclass[11pt, a4paper, logo, copyright]{googledeepmind}

\pdftrailerid{redacted}

\makeatletter
\renewcommand\bibentry[1]{\nocite{#1}{\frenchspacing\@nameuse{BR@r@#1\@extra@b@citeb}}}
\makeatother

\usepackage[utf8]{inputenc}
\usepackage{color}
\usepackage{graphicx}
\usepackage{wrapfig}
\usepackage{listings}
\newcolumntype{P}[1]{>{\centering\arraybackslash}p{#1}}

\newcommand{\commentout}[1]{}

\usepackage{subcaption}
\usepackage[font=footnotesize]{caption}
\usepackage[dvipsnames]{xcolor}%
\usepackage{upquote}
\usepackage{stackengine}
\newcommand*\colourcheck[1]{%
  \expandafter\newcommand\csname #1check\endcsname{\textcolor{#1}{\ding{51}}}%
}
\newcommand*\colourcross[1]{%
  \expandafter\newcommand\csname #1cross\endcsname{\textcolor{#1}{\ding{55}}}%
}

\colourcheck{YellowGreen}
\colourcross{Melon}

\usepackage{relsize}
\definecolor{themecolour}{rgb}{0.95, 0.95, 0.92}
\newcommand{\theme}[1]{\colorbox{themecolour}{\relsize{-1}\texttt{#1}}}

\definecolor{learnlmviolet}{HTML}{8A2BE2}

\usepackage{fancyvrb}

\usepackage{array}
\usepackage{collcell}
\newcommand{\parboxtop}[1]{\parbox[t]{\linewidth}{\raggedright #1}}
\newcolumntype{T}[1]{>{\collectcell\parboxtop}p{#1}<{\endcollectcell}}
\newcolumntype{L}[1]{>{\raggedright\arraybackslash}p{#1}}

\usepackage{placeins}
\usepackage{xspace}
\usepackage{hyperref}
\usepackage{kantlipsum, lipsum}
\usepackage{dsfont}
\usepackage{gdm-colors}
\usepackage{todonotes}
\usepackage{multirow}
\usepackage{comment}
\usepackage{spverbatim}
\usepackage{float} %
\usepackage{longtable}

\usepackage[numbers, sort&compress]{natbib}

\graphicspath{{figures/}}

\newcommand{\genai}{gen AI\xspace}

\newcolumntype{S}{>{\hsize=.5\hsize}X}

\title{LearnLM: Improving Gemini for Learning}

\correspondingauthor{learnlm-tech-report@google.com}

\paperurl{goo.gle/LearnLM-dec24}

\reportnumber{} %

\renewcommand{\today}{2024-12-19}

\author[]{LearnLM Team, Google}

\begin{abstract}
Today's generative AI systems are tuned to present information by default, rather than engage users in service of learning as a human tutor would. To address the wide range of potential education use cases for these systems, we reframe the challenge of injecting pedagogical behavior as one of \textit{pedagogical instruction following}, where training and evaluation examples include system-level instructions describing the specific pedagogy attributes present or desired in subsequent model turns. 
This framing avoids committing our models to any particular definition of pedagogy, and instead allows teachers or developers to specify desired model behavior.
It also clears a path to improving Gemini models for learning---by enabling the addition of our pedagogical data to post-training mixtures---alongside their rapidly expanding set of capabilities.
Both represent important changes from our initial tech report~\citep{jurenka2024towards}.
We show how training with pedagogical instruction following produces a LearnLM model (available on \href{http://goo.gle/LearnLMaccess}{Google AI Studio}) that experts substantially prefer across a diverse set of learning scenarios, with average preference strengths of +31\% over GPT-4o, +11\% over Claude 3.5 Sonnet, and +13\% over the Gemini~1.5~Pro model on which LearnLM was based.
\end{abstract}

\begin{document}

\maketitle

\section{Introduction}

Our \href{https://storage.googleapis.com/deepmind-media/LearnLM/LearnLM_paper.pdf}{initial tech report}~\citep{jurenka2024towards} from May 2024 surveyed the history and current landscape of education technology, discussed the potential impact of generative artificial intelligence (\genai) on education, and presented our collaborative approach to developing evaluations.

Following its publication, we received input from across the international education sector, including schools, educational technology (``EdTech'') companies, non-profit organizations, and government agencies eager to try our models or otherwise collaborate. Through review of these submissions, over 20 follow-up interviews, and input from Google product teams building \genai powered learning features, we can summarize the key findings as follows:
\begin{enumerate}
    \item Pedagogy\footnote{We use the term \textit{pedagogy} in as broad a sense as possible, certainly not limited to children, to evoke techniques of teaching and associated learning by humans.}, or rather, ideal behavior of an AI tutor, is prohibitively difficult to define given the wide range of grade-levels, subjects, languages, cultures, product designs, and philosophies that must be accommodated. While there are many commonalities, appropriate behavior in different contexts may be different or even contradictory,
    and it is best left to the developer or teacher to specify.
    \item In developing AI learning systems, the most commonly cited, immediately useful behavior in an underlying model is the ability to follow system instructions to create interactive tutor-led exercises. Teachers or developers who specify these instructions want to feel confident that the AI tutor will follow the specified instructions accurately, even if a student tries to circumvent them (e.g., ``do not give away the answer'' or ``stay on topic'').
    \item Post-hoc fine-tuning for each application can be effective in the short-term, but is impractical because of cost, maintenance, and rapidly improving base models. Thus, despite its shortcomings, prompting will likely remain the best way for education product developers to specify behavior.
\end{enumerate}

This paper describes how we have updated our modeling and evaluation methodology in light of these observations. Specifically, we cast our work as \textit{pedagogical instruction following}, meaning that we contextualize training and evaluation examples with system-level instructions that describe desired pedagogical behavior. This approach avoids any narrow specification of how systems should behave, and allows us to effectively add pedagogical data to the rest of Gemini's training mixture without conflicts of persona or style.
We also include Reinforcement Learning from Human Feedback (RLHF)~\cite{ziegler2019fine} in our training procedure to allow models to follow more nuanced pedagogical instructions and preferences.

\begin{figure}[t!]
    \centering
    \includegraphics[width=\textwidth]{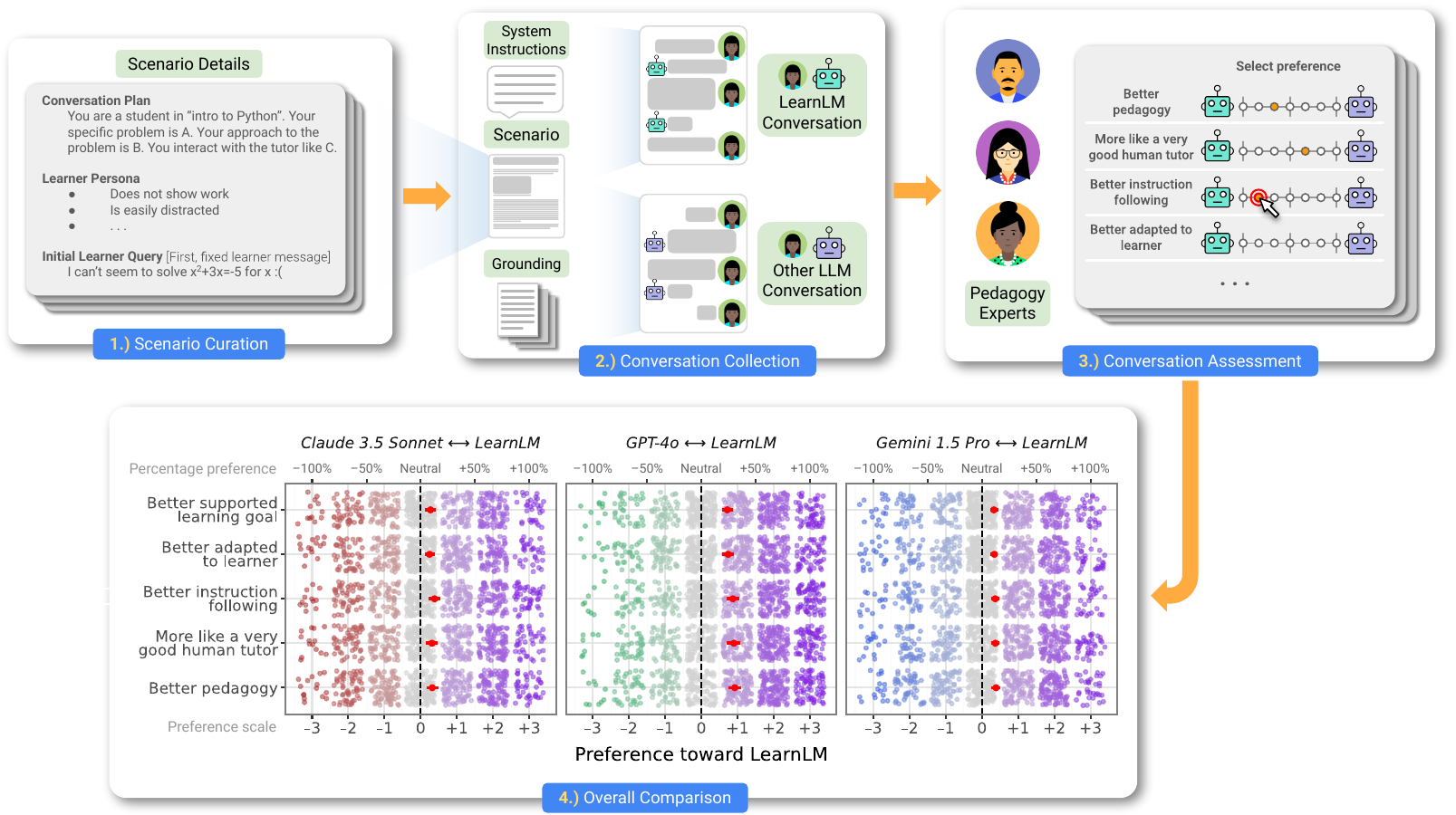}
\caption{An overview of our three-stage expert evaluation pipeline and our results for comparing LearnLM with other systems. (1) We developed learning scenarios that allow expert participants to role-play specific learners interacting with pairs of AI tutors. (2) Grounding material (e.g. an essay, homework problem, diagram, etc.) and System Instructions specific to each scenario are passed as context to each model. We pass the resulting conversation pairs to pedagogy experts (3) who review each model on its own, as well as their comparative performance. We aggregate these comparative assessments (on a seven-point response scale) (4) to evaluate overall preference for LearnLM over GPT-4o, Claude 3.5 Sonnet, and Gemini~1.5 Pro. See Section \ref{sec:results} for more detailed results.}
    \label{fig:high-level-overview}
\end{figure}

Using the updated methodology, we trained a new 
version of \emph{LearnLM},
based on Gemini 1.5 Pro\footnote{Specifically gemini-1.5-pro-002 (\href{https://cloud.google.com/vertex-ai/generative-ai/docs/learn/model-versions}{release notes}).}~\citep{geminiteam2024gemini}.
In our evaluations against contemporaneous flagship models, each representing a company’s premier offering as of 2024-10-01, 
educators and pedagogy experts preferred this version of LearnLM with an average preference strength of $+31\%$ over GPT-4o, $+11\%$ over Claude 3.5 Sonnet, and $+13\%$ over the original Gemini 1.5 Pro (see Figure \ref{fig:high-level-overview}).
LearnLM is available as an experimental model on \href{http://goo.gle/LearnLMaccess}{Google AI Studio} along with \href{https://ai.google.dev/gemini-api/docs/learnlm}{documentation} of example use cases and suggested prompts.
We welcome any \href{https://docs.google.com/forms/d/e/1FAIpQLSf5-B50OnNFjVGHLFkSerP1k0PZXHMgcnQ7k1cM_hIsqIjpjA/viewform}{feedback} on LearnLM to help inform our future research and priorities.
As we improve LearnLM for teaching and learning, we are also working to bring these advances into Gemini models, so any developers using Gemini can benefit from the improvements made via LearnLM research.

Section~\ref{sec:modeling} describes how we trained LearnLM for pedagogical instruction following and Section~\ref{sec:eval_design} explains how we updated our scenario-based evaluation design accordingly. 
Section \ref{sec:results} shows a detailed analysis of results comparing LearnLM with other premier model offerings.
Finally, Section \ref{sec:conclusion} outlines some future work, especially with regards to continued evaluation.
In addition to the broad range of core academic subjects that we use for training and evaluation, we include a feasibility study on medical education subjects in Appendix~\ref{sec:appendix/feasibility_study_on_medical_education_subjects}.

\section{Modeling}
\label{sec:modeling}

In our original tech report \cite{jurenka2024towards}, we adapted the behavior of a base model by Supervised Fine-Tuning (SFT) with a range of synthetic and human-written datasets. 
Since then, we have made a number of substantial changes to our training strategy:
First, we updated our SFT data according to our focus on pedagogical instruction following.
Second, we decided to additionally leverage Reinforcement Learning from Human Feedback (RLHF)\cite{ziegler2019fine}, for which we collected human preference data to train Reward Models (RMs) and prompts for the RL stage.
Third, rather than running our own post-training after Gemini's standard post-training, we \emph{co-train} with Gemini, meaning we mix our data directly with Gemini’s SFT, RM, and RL stages. LearnLM is the result of this experimental mixture and we have also been integrating our data and evaluations into the main Gemini models; a subset of LearnLM improvements is part of the recently released \href{https://blog.google/technology/google-deepmind/google-gemini-ai-update-december-2024/}{Gemini 2.0} models \cite{sundar2024geminiblog}.

\subsection{Pedagogical instruction following}

Instruction following (IF) refers to a model's ability to follow prompts, usually to better align with human intents \cite{ouyang2022training}. Gemini \cite{geminiteam2024gemini} differentiates between \textit{User Instructions}, inserted by a user during conversation, and \href{https://ai.google.dev/gemini-api/docs/system-instructions}{\textit{System Instructions}}, typically specified by a developer ahead of any user interaction, which take precedence over any subsequent instructions provided by the user. System Instructions can vary greatly in complexity, from a single minimally specified sentence like ``You are a knowledgeable writing coach'', to specific conditional expectations, e.g. ``If the user has answered 3 questions correctly, move to the next topic'', to detailed, multi-paragraph instructions that describe complex tasks and behaviors, exemplified by the education prompts in \citet{mollick2023assigning}, or the recently proposed Complex IF benchmark  \cite{wen2024benchmarking}.

Instructions broadly fall into two categories: hard constraints, often used for length, formatting, or content requirements (e.g. ``summarize the text in less than 100 words'' or ``do not use word X''), and soft, more nuanced constraints or guidelines, often used to control style, persona, or tone (e.g. ``use a professional voice'' or  ``use language that is easier to understand for a non-native speaker'').  Among open-source IF benchmarks, IFEval \cite{zhou2023instruction} focuses on programmatically verifiable IF, a subset of hard constraints, with more recent benchmarks like~\citet{qin2024infobench} expanding the scope to include more nuanced linguistic and stylistic guidelines. For educational use cases, both categories of instructions are important, e.g. ``do not reveal the answer'' is a hard constraint, while ``use a motivating tone'' is a soft one.

Improvements on IF capabilities have already resulted in better model responses for many learning use cases. In this work, we build on this progress and focus on improving instruction following for pedagogical System Instructions, which tend to be more complex, nuanced and not easily verifiable; these attributes make them more difficult for models to follow.

\subsection{Post-training and data collection strategy}

Our primary modeling strategy is to collect data that improves the models' ability to follow pedagogical System Instructions that we observed were common for developers building AI tutors. Accordingly, we updated our SFT data so that each conversation begins with a different System Instruction that specifically describes the pedagogical behavior present in that conversation. More general or vague instructions are counterproductive because the model learns to ignore instructions that are not useful for predicting the target model turns.

To collect human preference data, we similarly seed each conversation with a different pedagogically-focused System Instruction, and ask raters to label model samples based on the degree to which they adhere to those instructions. These conversations and turn-level labels are used to train a reward model, which is then employed during RLHF to score samples from the policy model. While SFT seems to improve pedagogical instruction following somewhat, RL is significantly more effective, as preference judgements often contain subtle distinctions in how instructions are interpreted and followed in the context of long conversations.

\subsection{Benefits of co-training}

Pedagogical behavior is often at odds with typical behavior of conversational AI, principally because learning is often a process of discovery rather than simply a transfer of information. Our instruction following approach allows us to mix pedagogical conversation data alongside data that contains more typical interactions by conditioning pedagogical model responses on specific System Instructions. By co-training with Gemini's post-training mixture, we allow the model to learn new kinds of instruction following without ``forgetting'' other core reasoning, multimodal understanding, factuality, safety, or multi-turn properties. Moving forward, we can also more easily keep LearnLM in sync with Gemini as the training recipe evolves.

\section{Expert Evaluation Design}
\label{sec:eval_design}

In our initial tech report, we discussed a taxonomy of pedagogy evaluation designs and reported results of four human evaluations with different methodologies (Sections 4 and 5 in \citet{jurenka2024towards}).
Within this taxonomy, here we focus on scenario-guided, conversation-level pedagogy evaluations and side-by-side comparisons.
For this new set of evaluations, we improved the clarity and coverage of our learning scenarios, added System Instructions specific to each scenario, and updated our pedagogy rubric and questions.
Guiding participant conversations with scenarios is especially important in multi-turn settings~\citep{ibrahim2024beyond}. Without scenarios, the unconstrained nature of human-AI interactions frequently leads to meandering conversations, offering a poor basis for comparison. In contrast, scenario-based approaches support relatively repeatable, controlled comparisons of the capabilities of different conversational AI systems. Scenario frameworks also help with evaluation coverage, ensuring that we test a diverse range of use cases. 

Our evaluation process takes place in three stages, depicted above in Figure~\ref{fig:high-level-overview}.
First, we identified an ecologically representative distribution of learning use cases and created a bank of 49 evaluation scenarios (Section~\ref{sec:eval_design/scenario_design}).
Second, these scenarios grounded interactions between AI systems and a pool of $N=186$ educators and pedagogy experts role-playing as learners across learning goals, subjects, learning materials, and learner personas (Section~\ref{sec:eval_design/conversation_collection}).
Third, to assess the quality of pedagogy in these interactions, we separately recruited a pool of $N=248$ educators and pedagogy experts to review the performance of the systems (Section~\ref{sec:eval_design/pedagogical_assessment}). This process produced ample quantitative and qualitative data to help us understand the systems' capabilities and behavior (Section~\ref{sec:eval_design/analysis}).

We are committed to following best practices in research ethics, including by communicating transparently about our research aims, collecting informed consent, and compensating fairly for participation~\citep{mckee2024human}. Our protocol underwent independent ethical review, with a favourable opinion from the Human Behavioural Research Ethics Committee at Google DeepMind (\#23 011).

\subsection{Scenario design} \label{sec:eval_design/scenario_design}

An \emph{evaluation scenario} is a structured template that supports consistent, multi-turn evaluations of conversational AI systems. A scenario specifies certain ``key properties'' about an interaction between an individual and an AI system, such as the goals, traits, and actions for the individual, as well as relevant conversational context. 
The scenarios that we curated ask human participants to role-play as different types of learners (e.g., students in classrooms, or independent EdTech users) across a wide range of learning contexts that vary by academic discipline, learning objective, and instructional approach. We used a systematic procedure to develop the bank of learning scenarios, drawing upon input from the educational ecosystem and support from pedagogy experts:

\paragraph{Phase 1: Use-case elicitation.} To begin the development of our scenario bank, we solicited feedback from EdTech companies, educational institutions, and Google product teams seeking to apply \genai to tutoring and teaching. We asked them to share common use cases, prompts, opportunities, and challenges they saw for \genai in real-world educational settings. We compiled and analyzed this feedback as a team with the goal of identifying common themes that should inform our evaluation approach.

\paragraph{Phase 2: Template design.} Based on these use cases, opportunities, and challenges, we drafted a structured scenario template (see ``Scenario structure and contents'' in Appendix~\ref{app:methods/scenario_contents}) and a specific protocol to steer scenario generation, including a set of guiding questions for each property (see ``Protocol for scenario generation'' in Appendix~\ref{app:methods/scenario_protocol}).

\paragraph{Phase 3: Scenario generation and refinement.} We next collaboratively and iteratively populated our bank of scenarios. Members of our team---including two with years of professional experience educating students and training teachers---independently drafted scenarios, leveraging the template and guiding questions from Phase 2. We collectively reviewed the scenario drafts, assessing each for clarity, completeness, correctness, and relevance to our pedagogical principles and the use cases defined in Phase 1. We weighted the overall distribution of scenarios across different learning goals, personas, and subject areas, flagging any gaps for further development.

This process resulted in a diverse bank of 49 scenarios across core academic subjects (for examples, see Appendix~\ref{app:methods/example_scenarios}).
In addition to building this foundational bank, we later validated the robustness and replicability of this procedure by running a feasibility study in a specialized educational domain (specifically, medical education; see Appendix~\ref{sec:appendix/feasibility_study_on_medical_education_subjects}).

\subsection{Conversation collection} \label{sec:eval_design/conversation_collection}
\begin{figure}[th!]
    \centering
    \stackinset{l}{1.6cm}{t}{0.6cm}{\hypertarget{fig:conversation_collection_instructions}{\phantom{A}}}{%
        \stackinset{l}{4.3cm}{t}{0cm}{\hypertarget{fig:conversation_collection_scenario}{\phantom{B}}}{%
            \stackinset{l}{12.4cm}{t}{0cm}{\hypertarget{fig:conversation_collection_configuration}{\phantom{C}}}{%
                \stackinset{l}{11cm}{t}{3.6cm}{\hypertarget{fig:conversation_collection_interaction}{\phantom{D}}}{%
                    \stackinset{l}{5.9cm}{t}{3.6cm}{\hypertarget{fig:conversation_collection_tutor_survey}{\phantom{E}}}{%
                        \stackinset{l}{0.4cm}{t}{3.6cm}{\hypertarget{fig:conversation_collection_comparison_survey}{\phantom{F}}}{%
                            \includegraphics[width=\linewidth]{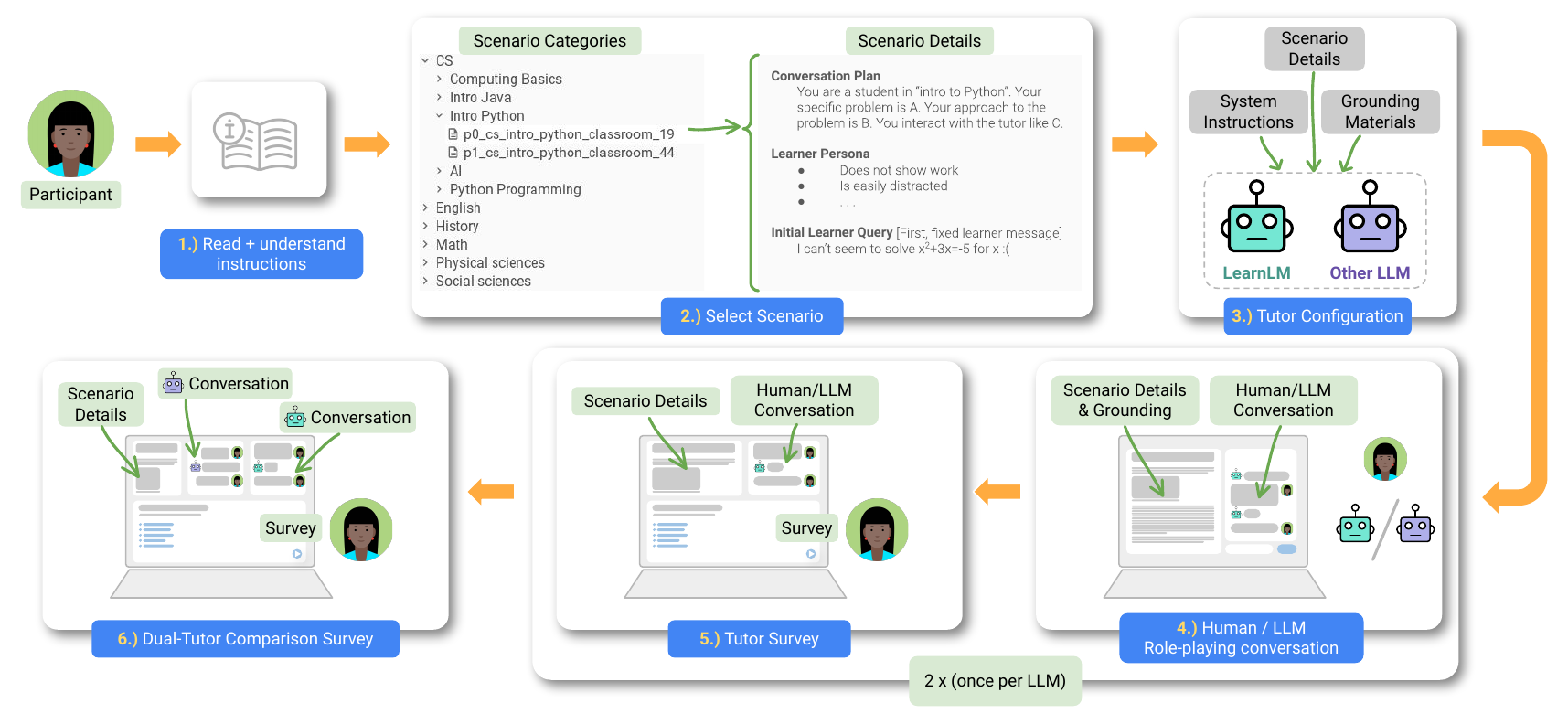}
    }}}}}}
    
    \caption{Workflow to generate conversations based on educational scenarios. A participant enacts conversations with prompted models as defined by scenarios. The participant then fills out a survey capturing quality and preference between models.}
    \label{fig:conversation_collection_ui_v2}
\end{figure}

In the second stage, we collected a corpus of conversations in which human participants role-played learners interacting with an AI system, as specified in the evaluation scenarios. To effectively simulate learner behavior in our educational scenarios, we recruited a pool of $N=168$ pedagogy experts with advanced academic degrees and two or more years of experience as a tutor.

Every session of conversation collection began with a short training on role-playing the scenarios (see Figure~\ref{fig:conversation_collection_ui_v2}, \hyperlink{fig:conversation_collection_instructions}{Step 1}). After passing a quiz at the end of the training, participants selected a scenario to enact (see Figure~\ref{fig:conversation_collection_ui_v2}, \hyperlink{fig:conversation_collection_scenario}{Step 2}). Conversation collection proceeded in pairs, such that the same participant enacted a scenario first with one AI system, and then another. Each pair included LearnLM and a comparison system. We randomized the order of the systems for each conversation pair and did not label the systems for participants. Within each pair of conversations, the models received the same System Instructions, grounding material, and initial learner queries as context, as specified by our scenarios (see Figure~\ref{fig:conversation_collection_ui_v2}, \hyperlink{fig:conversation_collection_configuration}{Step 3}). We formatted all inputs identically, except for some small specification differences mandated by the system APIs.

As specified by our template, each scenario included an initial query for the learner that was automatically sent on behalf of the participant to begin the conversation. After the AI system responded to that query, the participant continued the interaction, guided by the information provided in the scenario. We required participants to continue for a minimum of 10 conversational turns (thus, a minimum of five learner and five tutor turns) before they could end the interaction.

After ending each conversation (see Figure~\ref{fig:conversation_collection_ui_v2}, \hyperlink{fig:conversation_collection_interaction}{Step 4}), participants filled out a brief questionnaire to share their experience interacting with the system (see Figure~\ref{fig:conversation_collection_ui_v2}, \hyperlink{fig:conversation_collection_tutor_survey}{Step 5} \& Appendix~\ref{sec:appendix/conversation_collection_conversation_questions}). Additionally, after each pair of conversations, participants completed another questionnaire focused on comparing their impressions of the two systems (see Figure~\ref{fig:conversation_collection_ui_v2}, \hyperlink{fig:conversation_collection_comparison_survey}{Step 6} \& Appendix~\ref{sec:appendix/conversation_collection_comparative_questions}). Participants could then either select a new scenario to begin two additional conversations or end the session.

\subsection{Pedagogical assessment} \label{sec:eval_design/pedagogical_assessment}

Finally, in the third stage, we recruited another pool of $N=228$ pedagogy experts---again with advanced academic degrees and two or more years of experience as a tutor---to analyze these conversations and assess the pedagogical capabilities of the different AI models.

Each assessment session began with a short training on the goals of our evaluation and the scenario template. We randomly assigned each participant a scenario to review. After review, we randomly assigned them a pair of conversations from that scenario to assess (i.e., a pair of conversations collected from a single participant from the conversation-collection stage). Participants reviewed one conversation transcript at a time. After reading a transcript, participants answered a questionnaire focused on the pedagogical performance of the AI system from that conversation (see Appendix~\ref{sec:appendix/pedagogy_conversation_questions}). After every pair of conversations, participants completed an additional brief questionnaire comparing their assessment of the two systems (see Appendix~\ref{sec:appendix/pedagogy_comparative_questions}). We aimed to collect three independent assessments for each pair of conversations to reduce the effects of interrater variability.

\subsection{Analysis} \label{sec:eval_design/analysis}

We employ a Bayesian statistical framework for our quantitative analyses.
By directly quantifying the probability of hypotheses and providing a clear, interpretable measure of uncertainty, Bayesian analysis offers a practical, informative approach for evaluating AI systems intended for deployment in the real world.

Our study design involves repeated measurements from our participants. That is, each participant role-playing as a learner interacted with each system multiple times, and each expert assessed each system multiple times. To account for this non-independence and avoid artificially inflating our certainty in our estimates, we analyze our data with hierarchical regressions~\citep{gelman1995bayesian}.
Appendix~\ref{sec:appendix/bayesian} describes our statistical methods in more detail.

In addition, we conducted qualitative analysis of the open-ended comments and feedback collected from our experts after role-playing each scenario with two systems (Stage 2)\footnote{For our qualitative analysis, we prioritized the feedback from Stage 2 for its unique insights into the firsthand experience of learning with each tutor.}.
In particular, we identified and then refined general themes related to the learner-system interactions from participants' free-form responses. We then coded individual responses for the presence or absence of each theme. To avoid biasing our annotations, we censored the identities of the systems during this process.
Appendix~\ref{sec:appendix/codebook} presents the codebook that we developed through our analysis.
\section{Results}
\label{sec:results}
We compared LearnLM against contemporaneous flagship offerings (as of 2024-10-01), in particular
GPT-4o\footnote{GPT-4o version 2024-08-06, \url{https://platform.openai.com/docs/models/gpt-4o}.},
Claude 3.5 Sonnet\footnote{Claude 3.5 Sonnet version 2024-06-20,
\url{https://docs.anthropic.com/en/docs/about-claude/models}.},
and Gemini 1.5 Pro\footnote{Gemini 1.5 Pro-002 from 2024-09-24,
\url{https://cloud.google.com/vertex-ai/generative-ai/docs/learn/model-versions}.}. 
Since we undertook this specific evaluation, each of these models has been updated, with new versions released.
Our results should therefore be understood as a point-in-time comparison, evaluating the effectiveness of our approach and establishing a baseline for our continued investment in education.

\begin{figure}[h]
\centering
\hspace*{0.86cm}
\begin{tabular}{|l|c|c|c|}
\hline
\textbf{System} & \textbf{Version} & \textbf{Avg Turns per Conversation} & \textbf{Avg Words per Turn} \\ \hline
LearnLM & 2024-11-19 & 11.0 & 174 \\ \hline
Gemini 1.5 Pro & 2024-09-24 & 10.3 & 130 \\ \hline
GPT-4o & 2024-08-06 & 10.1 & 137 \\ \hline
Claude 3.5 Sonnet & 2024-06-20 & 9.7 & 179 \\ \hline
\end{tabular}
\includegraphics[width=\textwidth]{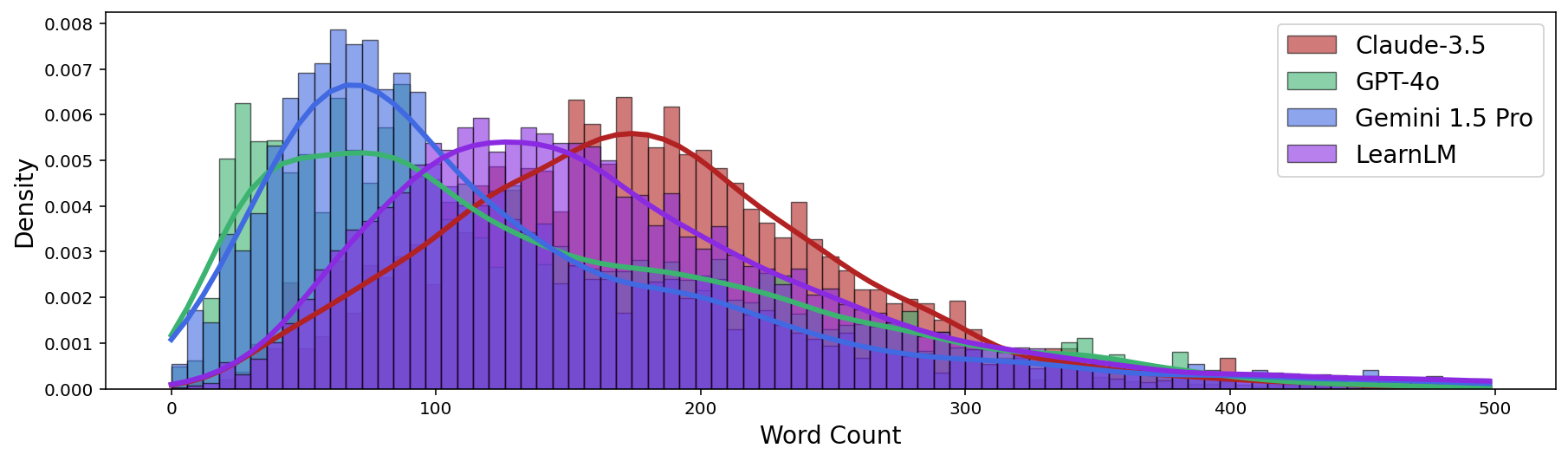}
\caption{(Top) The specific LLMs compared, along with aggregate statistics across all conversations collected: average number of model turns per conversation and average number of words per turn; (Bottom) Histograms of the number of words used per turn by each model.}
\label{fig:model-stats}
\end{figure}

In total, we collected a set of $2360$ conversations, consisting of $58\,459$ total learner and model messages. We collected $10\,192$ expert assessments of those conversations, with an average of three experts reviewing each pair of conversations.
Figure~\ref{fig:model-stats} shows that the systems we evaluate demonstrate notably different response length distributions across the collected conversations, including between Gemini 1.5 Pro and LearnLM. On an aggregate level, we observe no clear relationship between length and perceived quality (cf. \citep{stylearena2024}).

\begin{figure}[h!]
    \centering
    \includegraphics[width=\textwidth]{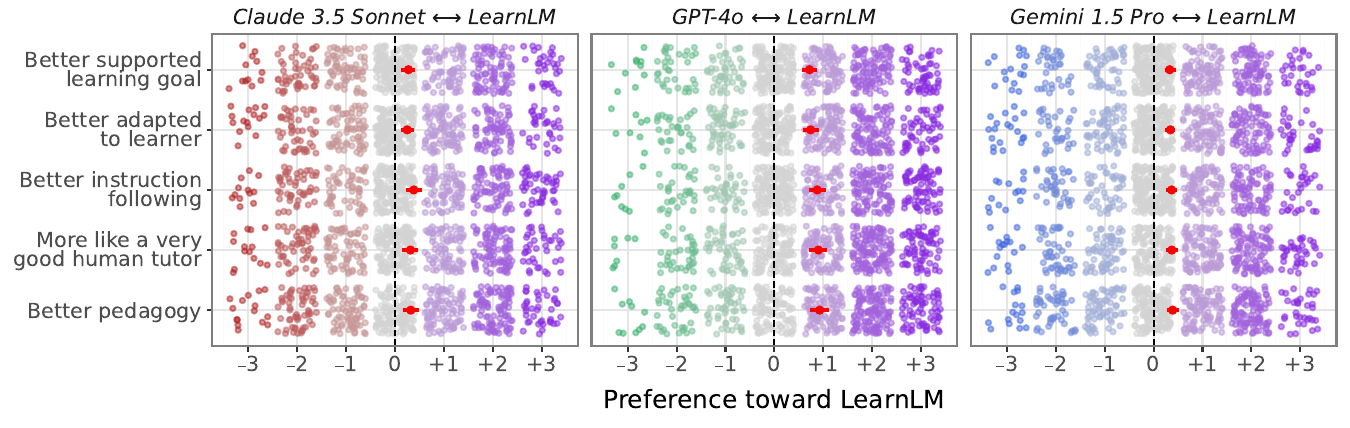}
    \caption{Pedagogy experts' preferences over LearnLM and other contemporaneous systems (Claude 3.5 Sonnet, GPT-4o, and Gemini 1.5 Pro). The scatterplots represent the underlying distribution of seven-point preference ratings. Given the large number of ratings we collected, these scatterplots proportionally downsample to 500 ratings per measure, color-coded based on the preference scale (dark purple corresponds to strong preference for LearnLM), and randomly jittered around each scale value for readability. The red points and error bars indicate the estimated mean and its 95\% credible interval for each measure.}
    \label{fig:sxs_results}
\end{figure}

We begin our analysis by examining the pedagogical assessments and preference ratings from our expert evaluators. We then explore the firsthand feedback from participants who role-played as learners interacting with the models. That is, we present our pedagogical findings from Stage 3 before backtracking to the interaction data gathered from Stage 2. Several clear patterns emerge from this analysis.

First, comparative preference ratings (Figure~\ref{fig:sxs_results}) reveal a strong preference toward LearnLM over GPT-4o for all five comparative assessment categories. Experts expressed the strongest preference for LearnLM in overall pedagogy (``Which tutor demonstrated better tutoring?''). They also communicated similar but smaller preferences toward LearnLM over Claude 3.5 Sonnet and Gemini 1.5 Pro. Because we adapted Gemini 1.5 Pro to train LearnLM, the comparison between those two models directly reflects the changes that result from adding pedagogical data (see Section~\ref{sec:modeling}).

\begin{figure}[h]
\centering
\includegraphics[width=\textwidth]{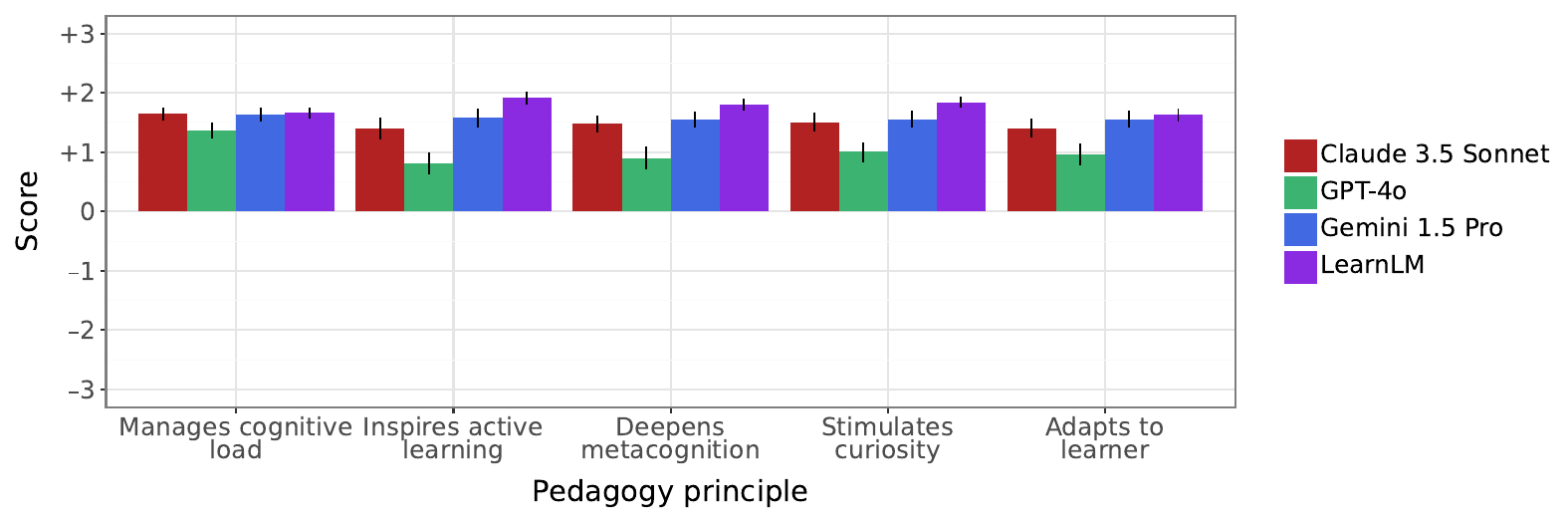}
\caption{Evaluation of systems in each category of our pedagogy rubric on a seven-point response scale (``Strongly disagree'' to ``Strongly agree''). Error bars reflect 95\% credible intervals from the posterior distrubtion for the mean.}
\label{fig:absolute_ratings}
\end{figure}

\begin{figure}
    \centering
    \includegraphics[height=5cm]{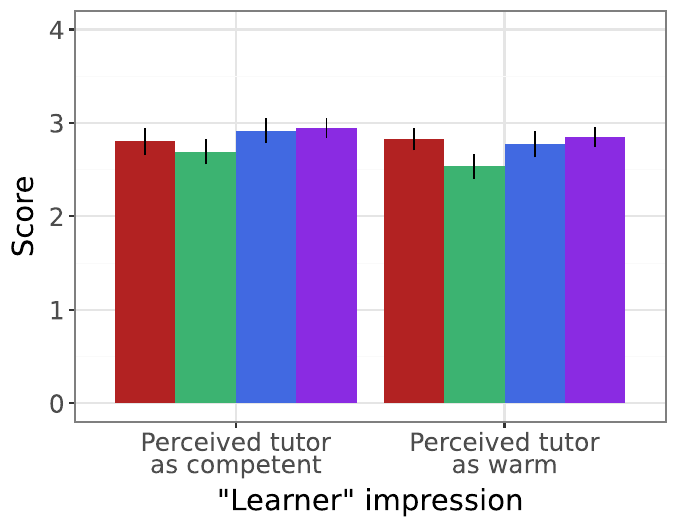} \vspace{1em} \includegraphics[height=5cm]{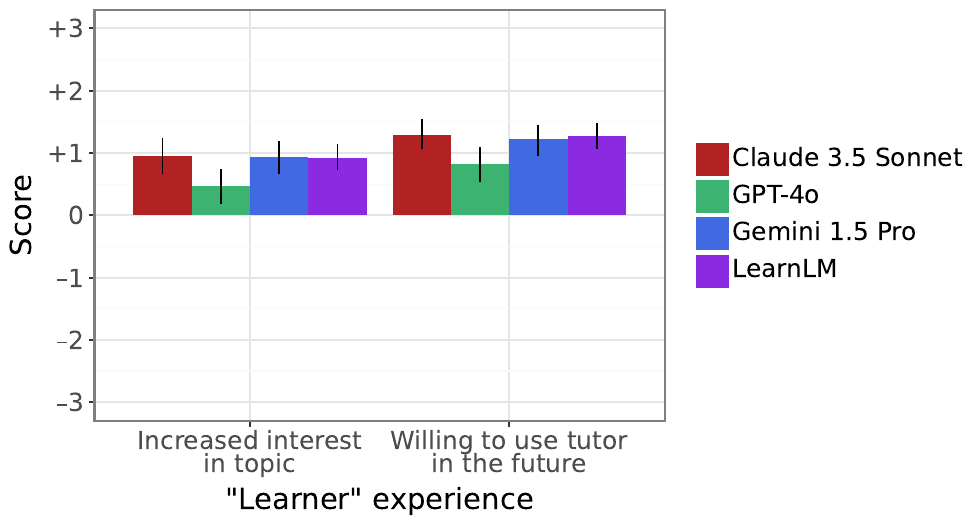}
    \caption{Impressions shared by the pedagogy experts role-playing as learners in our pedagogical scenarios. Error bars reflect 95\% credible intervals from the posterior distribution for the mean.
    Participants responded to the impression items on a five-point response scale (``Not at all'' to ``Extremely'') and to the experience items on a seven-point response scale (``Strongly disagree'' to ``Strongly agree'').
    }
    \label{fig:first_stage_ratings}
\end{figure}

Second, Figure \ref{fig:absolute_ratings} shows the mean performance of each model on our pedagogy rubric. Experts evaluated individual pedagogy qualities on a seven-point scale.
On average, each system received a positive assessment across every rubric category from this review.
Experts collectively assigned the highest marks to LearnLM across all rubric categories, and across almost all $29$ rubric questions, with particularly large leads on \textit{inspiring active learning}, \textit{deepening metacognition}, and \textit{stimulating curiosity}.

Third, Figure~\ref{fig:first_stage_ratings} depicts the degree to which each system increased participants' interest in the tutoring topic, participants' willingness to use the model in the future~\citep{davis1989perceived}, and their perceptions of the competence and warmth of the model~\citep{fiske2007universal, mckee2023humans}. 
Our participants reported relatively similar experiences with LearnLM, Gemini 1.5 Pro, and Claude 3.5 Sonnet. In contrast, participants indicated weaker experiences with GPT-4o in terms of its effects on their interest, its perceived warmth, and its perceived usefulness.
While role-playing experts represent imperfect proxies for students, their impressions help provide preliminary insight into the user experience of AI tutoring interactions.

Fourth, we randomly subsampled 203 explanations (approximately $20\%$~of the 1024 explanations that we collected) for thematic analysis of role-played learner preferences (see Table~\ref{tab:qualitative_coding_preference_explanations} for more details, including several example excerpts per theme). Overall, the themes that emerged most consistently in our subsample were \theme{is\_engaging} (appearing in $72$ of the subsampled explanations), \theme{conversation\_style} ($67$ explanations), and \theme{gives\_away\_answers} ($50$ explanations). 

When participants reported preferring LearnLM over the other model, their explanation was more likely to contain the themes \theme{keeps\_on\_topic}, \theme{challenges\_learner}, and \theme{gives\_away\_answers}. In contrast, participants who preferred other models to LearnLM tended to touch on the themes of \theme{clarity}, \theme{info\_amount}, and \theme{conversation\_style} in their explanations. Overall, our community of experts tended to see LearnLM as better at remaining on topic and guiding learners to a robust understanding of concepts, rather than simply giving away answers. On the other hand, these experts occasionally found LearnLM to be less suitable in terms of information delivery or conversation style.

\begin{table}[ht!]
\centering
\footnotesize    \begin{tabular}[t]{p{0.18\linewidth}L{0.135\linewidth}L{0.135\linewidth}T{0.44\linewidth}}
\toprule
\textbf{Theme}  & \textbf{Count when participants preferred LearnLM} (out of 94) & \textbf{Count when participants preferred other models} (out of 80) & \textbf{Example responses}\\
\midrule
\theme{keeps\_on\_topic} & $20~(\mathbf{21.2\%})$ & $8~(10\%)$ & \textcolor{learnlmviolet}{``[LearnLM] didn’t let me get away with distractions''} \newline
\textcolor{learnlmviolet}{``[LearnLM] was much more able to keep things on track''} \newline
\textit{``[The other tutor] also did a much better job of getting me back on task''}\\
\midrule
\theme{challenges\_learner} & $31~(\mathbf{33.0\%})$ & $13~(16.3\%)$ & \textcolor{learnlmviolet}{``obviously [LearnLM] was better [...] [the other tutor] clearly wasn’t pushing me to do well''} \newline
\textcolor{learnlmviolet}{``I felt like [LearnLM] was trying to help me grow and learn, rather than just agreeing with what I said''} \newline
\textit{``[The other tutor] asked interesting questions that made me think deeper''}
\\
\midrule
\theme{gives\_away\_answers} & $32~(\mathbf{34.0\%})$ & $15~(18.8\%)$ & \textcolor{learnlmviolet}{``[LearnLM] really engaged me in the steps to answer the question whereas [the other tutor] just gave me the answer''} \newline
\textcolor{learnlmviolet}{``[LearnLM] was keen on how to get the answer rather than giving the answer''}
\newline
\textit{``[LearnLM] was too reticent to help by giving answers when it was clear the student needed it''}\\
\midrule
\theme{clarity} & $15~(16.0\%)$ & $16~(\mathbf{20.0\%})$ & \textit{``The structure of the support [for the other tutor] was a bit clearer for the student to follow''} \newline
\textit{``[The other tutor] started smaller and simpler''} \newline
\textcolor{learnlmviolet}{``I just thought the answers [for LearnLM] were more clear''} \\
\midrule
\theme{info\_amount} & $19~(20.2\%)$ & $20~(\mathbf{25.0\%})$ & \textit{``[The other tutor] was [...] more succinct''} \newline
\textit{``[The other tutor] gave me everything I needed when I asked''}
\newline
\textcolor{learnlmviolet}{``[LearnLM] did a better job of breaking this "complex" topic into more digestible bites''}
\\
\midrule
\theme{conversation\_style} & $30~(31.9\%)$ & $29~(\mathbf{36.3\%})$ &
 \textit{``I [...] felt that [LearnLM] was a bit patronizing''} \newline
\textit{``[The other tutor] seemed warmer and more engaging''} \newline
\textcolor{learnlmviolet}{``[LearnLM] was warmer and more encouraging''}
\\
\bottomrule
\end{tabular}
\caption{Themes that were more likely to appear in ``learner'' explanations of preferences favoring LearnLM (top three rows), or favoring other models (bottom three rows). This table displays themes (i) referenced by at least $10\%$ of all sampled preference explanations, and (ii) showing an extreme ratio of occurrence between explanations favoring LearnLM and explanations favoring other models.}
\label{tab:qualitative_coding_preference_explanations}
\end{table}

\subsection{Safety evaluation} Similar to the process described in our initial tech report~\citep{jurenka2024towards} and the Gemini tech reports~\cite{geminiteam2024gemini, team2023gemini}, we carried out safety, responsibility, and assurance evaluations on LearnLM in collaboration with Google DeepMind's Responsible Development and Innovation team and Google's Trust and Safety team, with the goal of ensuring adherence to Gemini's model policy as well as a learning-specific model policy.

\paragraph{Model cards}
Our updated approach focuses on pedagogical instruction following and co-training with Gemini. Consequently, our training and safety evaluation procedure now fully aligns with that of Gemini 1.5. For its model card, see Table 45, Appendix 12 of its report~\cite{geminiteam2024gemini}. As a reminder, we describe our modeling approach, including learning-specific dataset curation, in Section~\ref{sec:modeling}. Our initial tech report~\cite{jurenka2024towards} provides the original model card for LearnLM, as well as a broader discussion of ethical risks and limitations for this area of research.
\section{Conclusion}
\label{sec:conclusion}

We have described our motivation and approach to improving foundation models for learning use cases, which relies on System Instructions to condition desired behavior. We updated Gemini's post-training mixture to add demonstration data (via SFT) and human preference data (via a Reward Model and RLHF) to teach the model to follow a range of pedagogical instructions. We then evaluated the resulting LearnLM model alongside comparable models, showing substantial preference for LearnLM, especially in instruction following capability, and more broadly across many pedagogical dimensions. The work described here represents the beginning of our effort to improve Gemini for learning use cases, as we bring the advances from LearnLM into Gemini\footnote{At the time of publication, some of our data has already been added to Gemini 2 models~\citep{sundar2024geminiblog}.}. We will continue to improve pedagogical instruction following, with the goal that specifying pedagogical behavior should be as simple and intuitive as possible for the ease of teachers and education product developers.

In addition to model improvements, we are planning more updates to our evaluation methodology. First, we want to work toward more consensus on a universal framework for pedagogical assessment of AI systems.
Although learning science principles underlie our current pedagogy rubric (see Appendix~\ref{sec:appendix/pedagogy_comparative_questions}), we need to work more closely with a diverse set of stakeholders to make sure it is appropriate for all learners and achieves the trust and approval of the broader education community.

Second, we would like to start moving from intrinsic evaluations,
which measure the model's performance according to a predefined pedagogy standard, 
to extrinsic evaluation,
which measure impact such as learning outcomes.
Intrinsic evaluations are useful for model development, as they are faster to run and directly identify the shortcomings in the models.
However, while the core principles of our rubric, such as encouraging active learning and managing cognitive load, are broadly agreed upon and evidence-based \cite{kirschner2020learning}, it is unclear how well the results translate to improvements in learning outcomes.
It is likely that as the field matures and AI systems master the basics of tutoring dialogue, extrinsic evaluations will play a more important role.
Recently, they have been used both for demonstrating improvements in learning outcomes \cite{kestin2024ai, wang2024tutor} and for comparing different systems and prompts \cite{bastani2024generative}.

Finally, we have begun to explore evaluations beyond core academic subjects. Our initial feasibility study in medical education (Appendix~\ref{sec:appendix/feasibility_study_on_medical_education_subjects}) confirms that our approach can generalize effectively to specialized domains.
As we continue to improve Gemini for use across a diverse range of educational settings, we welcome insights from applications of LearnLM to help us work towards realizing the potential of AI in education and learning \cite{nea2024, lomis2021artificial, desai2023precision}.

\bibliographystyle{unsrtnat}
\nobibliography*
\bibliography{main}

\begin{thebibliography}{25}
\providecommand{\natexlab}[1]{#1}
\providecommand{\url}[1]{\texttt{#1}}
\expandafter\ifx\csname urlstyle\endcsname\relax
  \providecommand{\doi}[1]{doi: #1}\else
  \providecommand{\doi}{doi: \begingroup \urlstyle{rm}\Url}\fi

\bibitem[Jurenka et~al.(2024)Jurenka, Kunesch, McKee, Gillick, Zhu, Wiltberger,
  Phal, Hermann, Kasenberg, Bhoopchand, et~al.]{jurenka2024towards}
Irina Jurenka, Markus Kunesch, Kevin~R McKee, Daniel Gillick, Shaojian Zhu,
  Sara Wiltberger, Shubham~Milind Phal, Katherine Hermann, Daniel Kasenberg,
  Avishkar Bhoopchand, et~al.
\newblock Towards responsible development of generative ai for education: An
  evaluation-driven approach.
\newblock \emph{arXiv preprint arXiv:2407.12687}, 2024.

\bibitem[Ziegler et~al.(2019)Ziegler, Stiennon, Wu, Brown, Radford, Amodei,
  Christiano, and Irving]{ziegler2019fine}
Daniel~M Ziegler, Nisan Stiennon, Jeffrey Wu, Tom~B Brown, Alec Radford, Dario
  Amodei, Paul Christiano, and Geoffrey Irving.
\newblock Fine-tuning language models from human preferences.
\newblock \emph{arXiv preprint arXiv:1909.08593}, 2019.

\bibitem[Team et~al.(2024)Team, Georgiev, Lei, Burnell, Bai, Gulati, Tanzer,
  Vincent, Pan, Wang, et~al.]{geminiteam2024gemini}
Gemini Team, Petko Georgiev, Ving~Ian Lei, Ryan Burnell, Libin Bai, Anmol
  Gulati, Garrett Tanzer, Damien Vincent, Zhufeng Pan, Shibo Wang, et~al.
\newblock Gemini 1.5: Unlocking multimodal understanding across millions of
  tokens of context.
\newblock \emph{arXiv preprint arXiv:2403.05530}, 2024.

\bibitem[Pichai et~al.(2024)Pichai, Hassabis, and
  Kavukcuoglu]{sundar2024geminiblog}
Sundar Pichai, Demis Hassabis, and Koray Kavukcuoglu.
\newblock Introducing gemini 2.0: our new ai model for the agentic era.
\newblock
  \url{https://blog.google/technology/google-deepmind/google-gemini-ai-update-december-2024/},
  2024.

\bibitem[Ouyang et~al.(2022)Ouyang, Wu, Jiang, Almeida, Wainwright, Mishkin,
  Zhang, Agarwal, Slama, Ray, et~al.]{ouyang2022training}
Long Ouyang, Jeffrey Wu, Xu~Jiang, Diogo Almeida, Carroll Wainwright, Pamela
  Mishkin, Chong Zhang, Sandhini Agarwal, Katarina Slama, Alex Ray, et~al.
\newblock Training language models to follow instructions with human feedback.
\newblock \emph{Advances in neural information processing systems},
  35:\penalty0 27730--27744, 2022.

\bibitem[Mollick and Mollick(2023)]{mollick2023assigning}
Ethan Mollick and Lilach Mollick.
\newblock Assigning ai: Seven approaches for students, with prompts.
\newblock \emph{arXiv preprint arXiv:2306.10052}, 2023.

\bibitem[Wen et~al.(2024)Wen, Ke, Gu, Wu, Huang, Zhou, Li, Hu, Gao, Xu,
  et~al.]{wen2024benchmarking}
Bosi Wen, Pei Ke, Xiaotao Gu, Lindong Wu, Hao Huang, Jinfeng Zhou, Wenchuang
  Li, Binxin Hu, Wendy Gao, Jiaxin Xu, et~al.
\newblock Benchmarking complex instruction-following with multiple constraints
  composition.
\newblock \emph{arXiv preprint arXiv:2407.03978}, 2024.

\bibitem[Zhou et~al.(2023)Zhou, Lu, Mishra, Brahma, Basu, Luan, Zhou, and
  Hou]{zhou2023instruction}
Jeffrey Zhou, Tianjian Lu, Swaroop Mishra, Siddhartha Brahma, Sujoy Basu,
  Yi~Luan, Denny Zhou, and Le~Hou.
\newblock Instruction-following evaluation for large language models.
\newblock \emph{arXiv preprint arXiv:2311.07911}, 2023.

\bibitem[Qin et~al.(2024)Qin, Song, Hu, Yao, Cho, Wang, Wu, Liu, Liu, and
  Yu]{qin2024infobench}
Yiwei Qin, Kaiqiang Song, Yebowen Hu, Wenlin Yao, Sangwoo Cho, Xiaoyang Wang,
  Xuansheng Wu, Fei Liu, Pengfei Liu, and Dong Yu.
\newblock Infobench: Evaluating instruction following ability in large language
  models.
\newblock \emph{arXiv preprint arXiv:2401.03601}, 2024.

\bibitem[Ibrahim et~al.(2024)Ibrahim, Huang, Ahmad, and
  Anderljung]{ibrahim2024beyond}
Lujain Ibrahim, Saffron Huang, Lama Ahmad, and Markus Anderljung.
\newblock Beyond static ai evaluations: advancing human interaction evaluations
  for llm harms and risks.
\newblock \emph{arXiv preprint arXiv:2405.10632}, 2024.

\bibitem[McKee(2024)]{mckee2024human}
Kevin~R McKee.
\newblock Human participants in {AI} research: Ethics and transparency in
  practice.
\newblock \emph{IEEE Transactions on Technology and Society}, 5\penalty0
  (3):\penalty0 279--288, 2024.
\newblock \doi{10.1109/TTS.2024.3446183}.

\bibitem[Gelman et~al.(1995)Gelman, Carlin, Stern, and
  Rubin]{gelman1995bayesian}
Andrew Gelman, John~B Carlin, Hal~S Stern, and Donald~B Rubin.
\newblock \emph{Bayesian data analysis}.
\newblock Chapman and Hall/CRC, 1995.

\bibitem[Tianle~Li(2024)]{stylearena2024}
Wei-Lin~Chiang Tianle~Li, Anastasios~Angelopoulos.
\newblock Does style matter? disentangling style and substance in chatbot
  arena.
\newblock \url{https://blog.lmarena.ai/blog/2024/style-control/}, August 2024.

\bibitem[Davis(1989)]{davis1989perceived}
Fred~D Davis.
\newblock Perceived usefulness, perceived ease of use and user acceptance of
  information technology.
\newblock \emph{MIS quarterly}, 1989.

\bibitem[Fiske et~al.(2007)Fiske, Cuddy, and Glick]{fiske2007universal}
Susan~T Fiske, Amy~JC Cuddy, and Peter Glick.
\newblock Universal dimensions of social cognition: Warmth and competence.
\newblock \emph{Trends in cognitive sciences}, 11\penalty0 (2):\penalty0
  77--83, 2007.

\bibitem[McKee et~al.(2023)McKee, Bai, and Fiske]{mckee2023humans}
Kevin~R McKee, Xuechunzi Bai, and Susan~T Fiske.
\newblock Humans perceive warmth and competence in artificial intelligence.
\newblock \emph{iScience}, 26\penalty0 (8), 2023.
\newblock \doi{10.1016/j.isci.2023.107256}.

\bibitem[Gemini et~al.(2023)Gemini, Anil, Borgeaud, Wu, Alayrac, Yu, Soricut,
  Schalkwyk, Dai, Hauth, et~al.]{team2023gemini}
Team Gemini, Rohan Anil, Sebastian Borgeaud, Yonghui Wu, Jean-Baptiste Alayrac,
  Jiahui Yu, Radu Soricut, Johan Schalkwyk, Andrew~M Dai, Anja Hauth, et~al.
\newblock Gemini: A family of highly capable multimodal models.
\newblock \emph{arXiv preprint arXiv:2312.11805}, 2023.

\bibitem[Kirschner and Hendrick(2020)]{kirschner2020learning}
Paul~A Kirschner and Carl Hendrick.
\newblock \emph{How learning happens: Seminal works in educational psychology
  and what they mean in practice}.
\newblock Routledge, 2020.

\bibitem[Kestin et~al.(2024)Kestin, Miller, Klales, Milbourne, and
  Ponti]{kestin2024ai}
Gregory Kestin, Kelly Miller, Anna Klales, Timothy Milbourne, and Gregorio
  Ponti.
\newblock Ai tutoring outperforms active learning.
\newblock 2024.

\bibitem[Wang et~al.(2024)Wang, Ribeiro, Robinson, Loeb, and
  Demszky]{wang2024tutor}
Rose~E Wang, Ana~T Ribeiro, Carly~D Robinson, Susanna Loeb, and Dora Demszky.
\newblock Tutor copilot: A human-ai approach for scaling real-time expertise.
\newblock \emph{arXiv preprint arXiv:2410.03017}, 2024.

\bibitem[Bastani et~al.(2024)Bastani, Bastani, Sungu, Ge, Kabakc{\i}, and
  Mariman]{bastani2024generative}
Hamsa Bastani, Osbert Bastani, Alp Sungu, Haosen Ge, Ozge Kabakc{\i}, and Rei
  Mariman.
\newblock Generative ai can harm learning.
\newblock \emph{Available at SSRN}, 4895486, 2024.

\bibitem[{National Education Association (NEA)}()]{nea2024}
{National Education Association (NEA)}.
\newblock Teaching and learning in the age of artificial intelligence.
\newblock
  \url{https://www.nea.org/resource-library/artificial-intelligence-education/iv-teaching-and-learning-age-artificial-intelligence}.
\newblock Accessed: 2024-12-10.

\bibitem[Lomis et~al.(2021)Lomis, Jeffries, Palatta, Sage, Sheikh, Sheperis,
  and Whelan]{lomis2021artificial}
Kimberly Lomis, Pamela Jeffries, Anthony Palatta, Melanie Sage, Javaid Sheikh,
  Carl Sheperis, and Alison Whelan.
\newblock Artificial intelligence for health professions educators.
\newblock \emph{NAM perspectives}, 2021, 2021.

\bibitem[Desai et~al.(2023)Desai, Burk-Rafel, Lomis, Caverzagie, Richardson,
  O’Brien, Andrews, Heckman, Henderson, Prober, et~al.]{desai2023precision}
Sanjay~V Desai, Jesse Burk-Rafel, Kimberly~D Lomis, Kelly Caverzagie, Judee
  Richardson, Celia~Laird O’Brien, John Andrews, Kevin Heckman, David
  Henderson, Charles~G Prober, et~al.
\newblock Precision education: the future of lifelong learning in medicine.
\newblock \emph{Academic Medicine}, pages 10--1097, 2023.

\bibitem[McKee et~al.(2024)McKee, Bai, and Fiske]{mckee2024warmth}
Kevin~R McKee, Xuechunzi Bai, and Susan~T Fiske.
\newblock Warmth and competence in human-agent cooperation.
\newblock \emph{Autonomous Agents and Multi-Agent Systems}, 38\penalty0
  (1):\penalty0 23, 2024.

\end{thebibliography}

\section*{Contributions and Acknowledgments}
\paragraph{Core Contributors}
Abhinit Modi,
Aditya Srikanth Veerubhotla,
Aliya Rysbek,
Andrea Huber,
Brett Wiltshire,
Brian Veprek,
Daniel Gillick,
Daniel Kasenberg,
Derek Ahmed,
Irina Jurenka,
James Cohan,
Jennifer She,
Julia Wilkowski,
Kaiz Alarakyia,
Kevin R. McKee,
Lisa Wang,
Markus Kunesch,
Mike Schaekermann,
Miruna Pîslar,
Nikhil Joshi,
Parsa Mahmoudieh,
Paul Jhun,
Sara Wiltberger,
Shakir Mohamed,
Shashank Agarwal,
Shubham Milind Phal,
Sun Jae Lee,
Theofilos Strinopoulos,
Wei-Jen Ko.

\paragraph{Contributors}
Amy Wang,
Ankit Anand,
Avishkar Bhoopchand,
Dan Wild,
Divya Pandya,
Filip Bar,
Garth Graham,
Holger Winnemoeller,
Mahvish Nagda,
Prateek Kolhar,
Renee Schneider,
Shaojian Zhu,
Stephanie Chan,
Steve Yadlowsky,
Viknesh Sounderajah,
Yannis Assael.

The roles are defined as follows:
\emph{Core Contributors} had direct and significant impact on the work presented in this report.
\emph{Contributors} made contributions to the work presented in this report.
Within each role, the order is alphabetical and does not indicate ordering of contributions.

\subsection*{Acknowledgements}
This work was done as part of the LearnLM effort, which is a cross-Google project, with members
from Google DeepMind (GDM), Google Research (GR), Google LearnX, Google Health, Google Creative Lab, YouTube Learning, YouTube Health, and more.
This tech report---focused on improvements to pedagogical instruction following---only represents a small part of the wider effort and only direct contributions are included in the contributor lists above.

Our work is made possible by the dedication and efforts of numerous teams at Google. We would like to acknowledge the support from:
Ajay Kannan,
Anand Rao,
Anisha Choudhury,
April (Soler) Manos,
Dawn Chen,
Dharti Dhami,
Edward Grefenstette,
Gal Elidan,
Himanshu Kattelu,
Jaume Sanchez Elias,
Jiao Sun,
Josh Capilouto,
Jyoti Gupta,
Kalpesh Krishna,
Lauren Winer,
Mac McAllister,
Mana Jabbour,
Michael Howell,
Miriam Schneider,
Muktha Ananda,
Nir Levine,
Niv Efron,
Ryan Muller,
Safwan Choudhury,
Shyam Upadhyay,
Svetlana Grant,
Tejasi Latkar,
William Wong,
Yael Haramaty.
Furthermore, we would like to thank Google DeepMind's Gemini team, Google DeepMind’s Responsible Development and Innovation, Responsible Engineering, and Child Safety teams and Google’s Trust and Safety team.
Finally, we would like to acknowledge the support from all our leads and sponsors to make this project happen.

\appendix
\section{Additional results}

\subsection{Preferences for participants role-playing as learners} 

\begin{figure}[h]
    \centering
    \includegraphics[width=\textwidth]{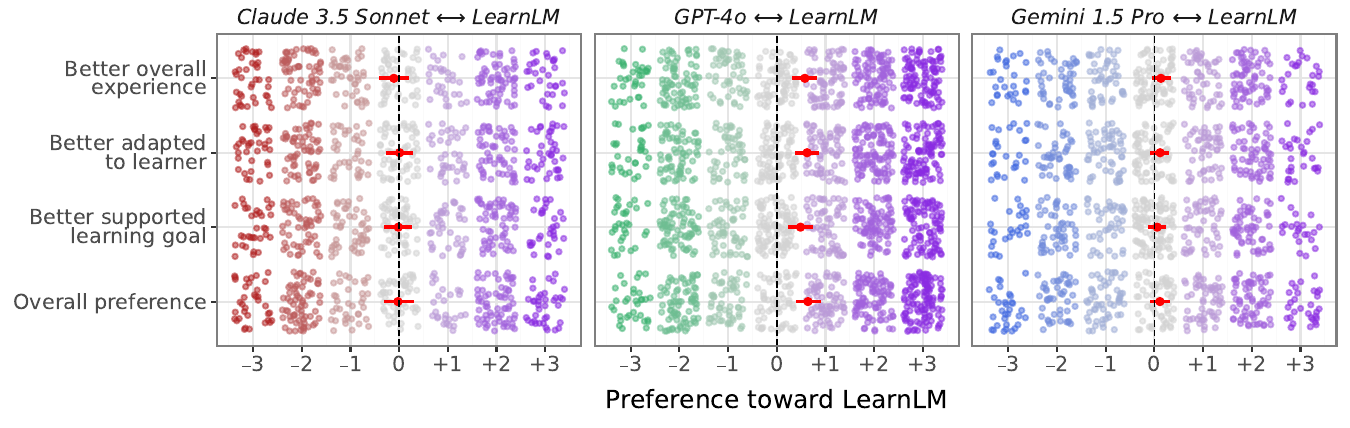}
    \caption{Preferences over LearnLM and other contemporary models (Claude 3.5 Sonnet, GPT-4o, and Gemini 1.5 Pro) according to the pedagogical experts role-playing as learners. The scatterplots represent the underlying distribution of seven-point preference ratings. Given the large number of ratings we collected, these scatterplots proportionally downsample to 500 ratings per measure. The red points and error bars indicate the estimated mean and its 95\% credible interval for each measure.}
    \label{fig:sxs_student_results}
\end{figure}

The participants role-playing as learners revealed a preference toward LearnLM over GPT-4o for all four comparative assessment categories  (Figure~\ref{fig:sxs_student_results}). Experts expressed the strongest preference for LearnLM in overall pedagogy (``Which tutor demonstrated better tutoring?'') and in similarity to a quality human tutor (``Which tutor was more like a very good human tutor?''). These participants indicated no substantial preference between LearnLM and Gemini 1.5 Pro or between LearnLM and Claude 3.5 Sonnet.

\subsection{Learner quality in collected conversations}

\begin{figure}[h]
    \centering
    \includegraphics[width=0.6\textwidth]{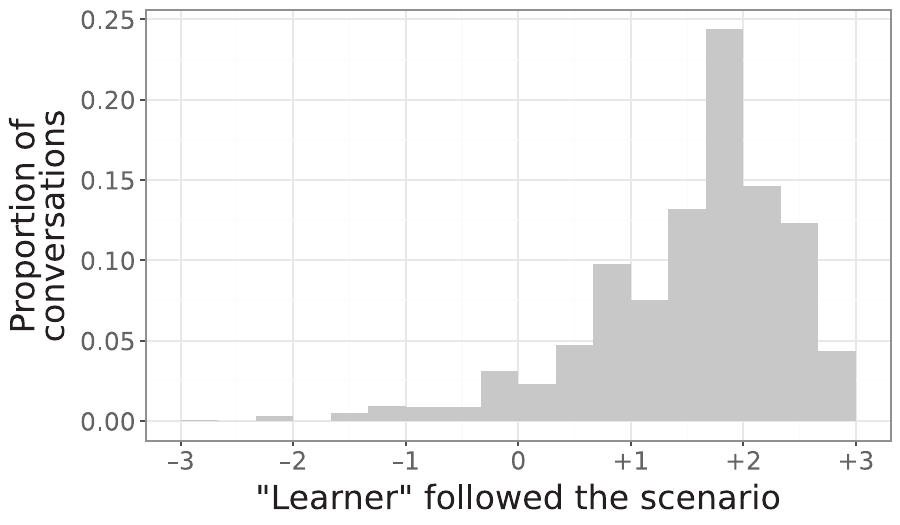}
    \caption{At the beginning of the pedagogical assessment process, we asked experts to evaluate how closely the human participants in the conversation transcripts followed the scenario instructions (i.e., how effectively they role-played the learner in the scenario) on a seven-point scale. This plot shows the responses grouped and averaged by transcript. These aggregate ratings indicate that the ``learner'' followed the scenario instructions in 93.2\% of conversation transcripts.}
    \label{app/fig:student_follows_persona}
\end{figure}

\clearpage
\subsection{Pedagogical assessment: detailed results}
\vspace*{\fill}

\begin{figure}[h]
    \centering
    \includegraphics[width=0.9\textwidth]{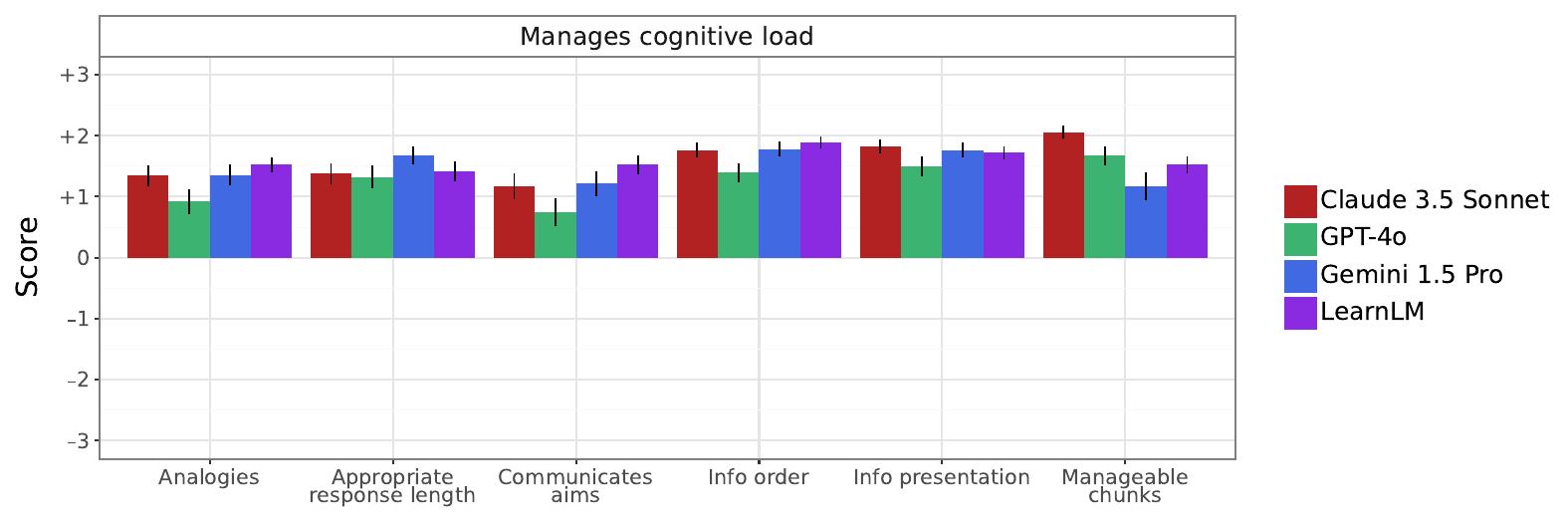} \\
    \includegraphics[width=0.9\textwidth]{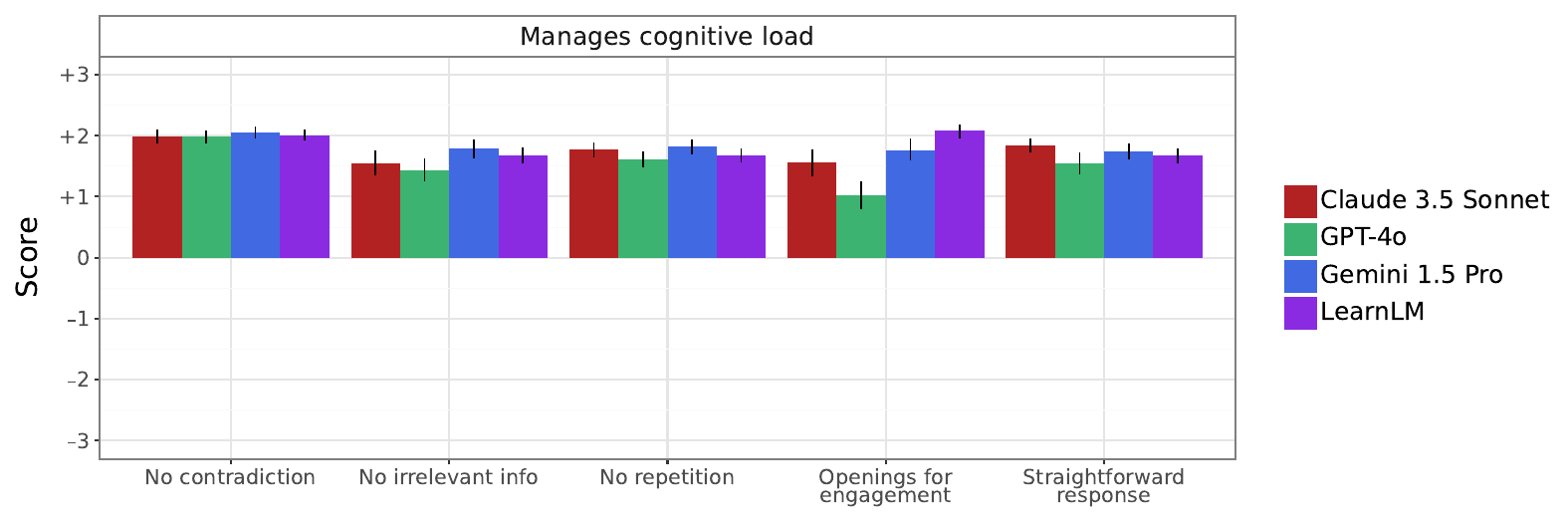}
    \caption{Evaluation of tutor models on specific subdimensions of the ``Cognitive load'' rubric category. Error bars reflect 95\% credible intervals from the posterior distribution for the mean.}
    \label{app/fig:cognitive_load}
\end{figure}

\vspace*{\fill}

\begin{figure}[h]
    \centering
    \includegraphics[width=0.9\textwidth]{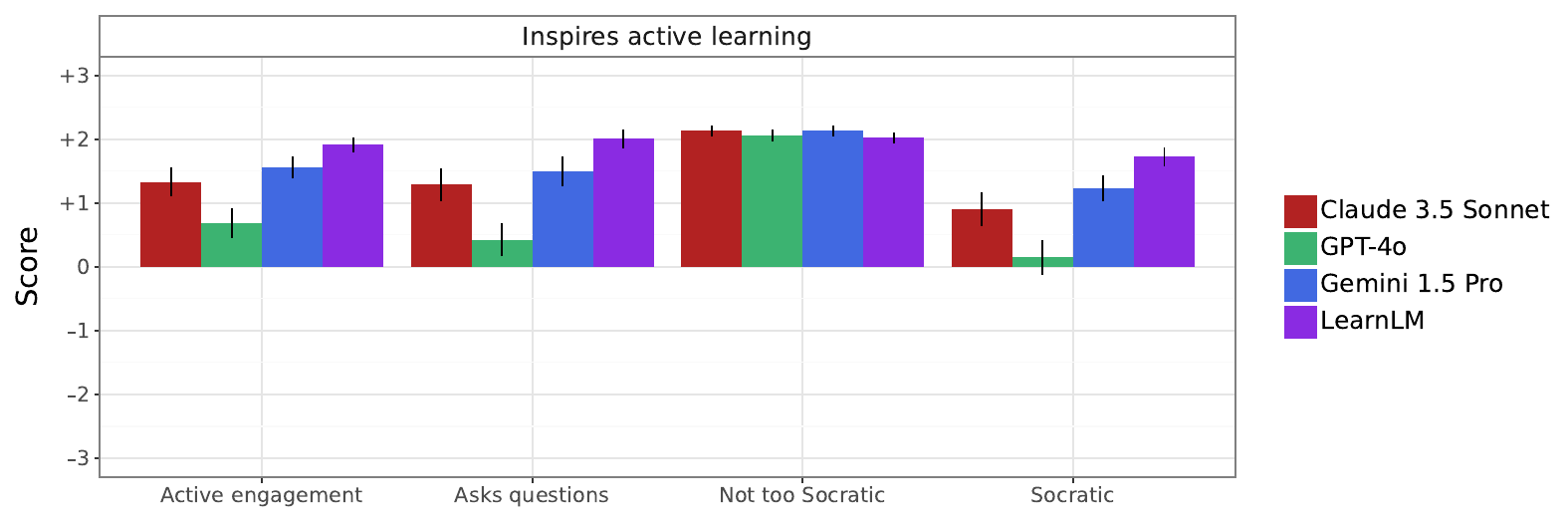}
    \caption{Evaluation of tutor models on specific subdimensions of the ``Active learning'' rubric category. Error bars reflect 95\% credible intervals from the posterior distribution for the mean.}
    \label{app/fig:active_learning_questions}
\end{figure}

\vspace*{\fill}
\newpage
\vspace*{\fill}

\begin{figure}[!ht]
    \centering
    \includegraphics[width=0.9\textwidth]{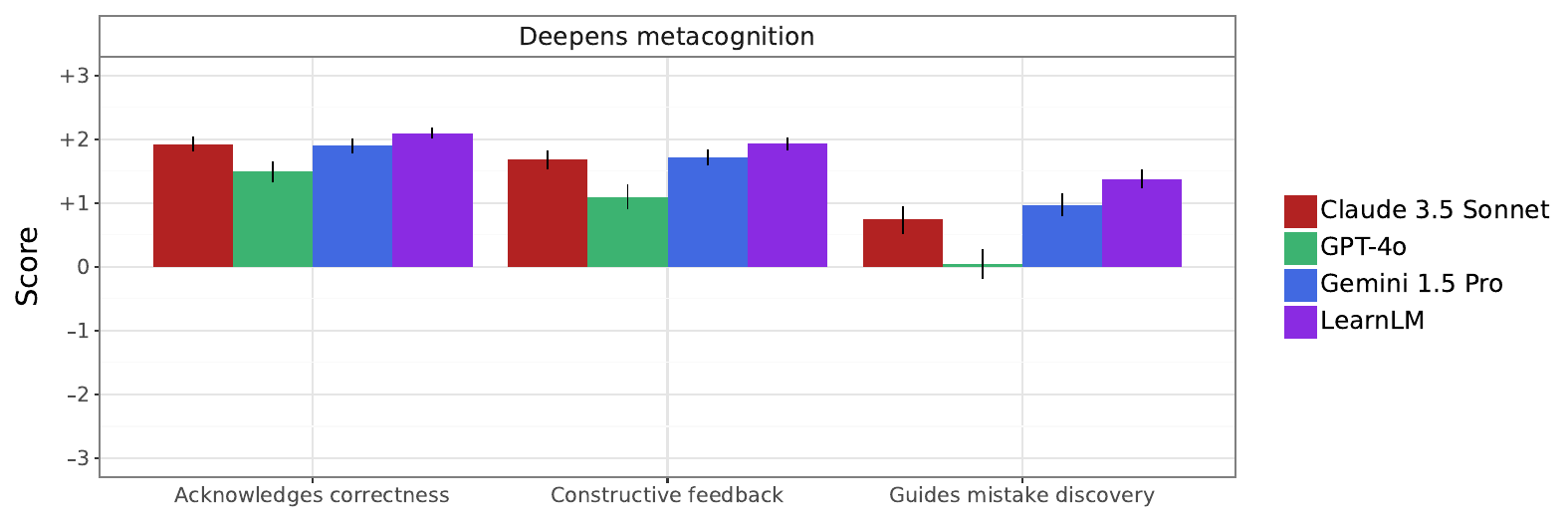}
    \caption{Evaluation of tutor models on specific subdimensions of the ``Deepen metacognition'' rubric category. Error bars reflect 95\% credible intervals from the posterior distribution for the mean.}
    \label{app/fig:deepen_metacognition}
\end{figure}

\vspace*{\fill}

\begin{figure}[!ht]
    \centering
    \includegraphics[width=0.9\textwidth]{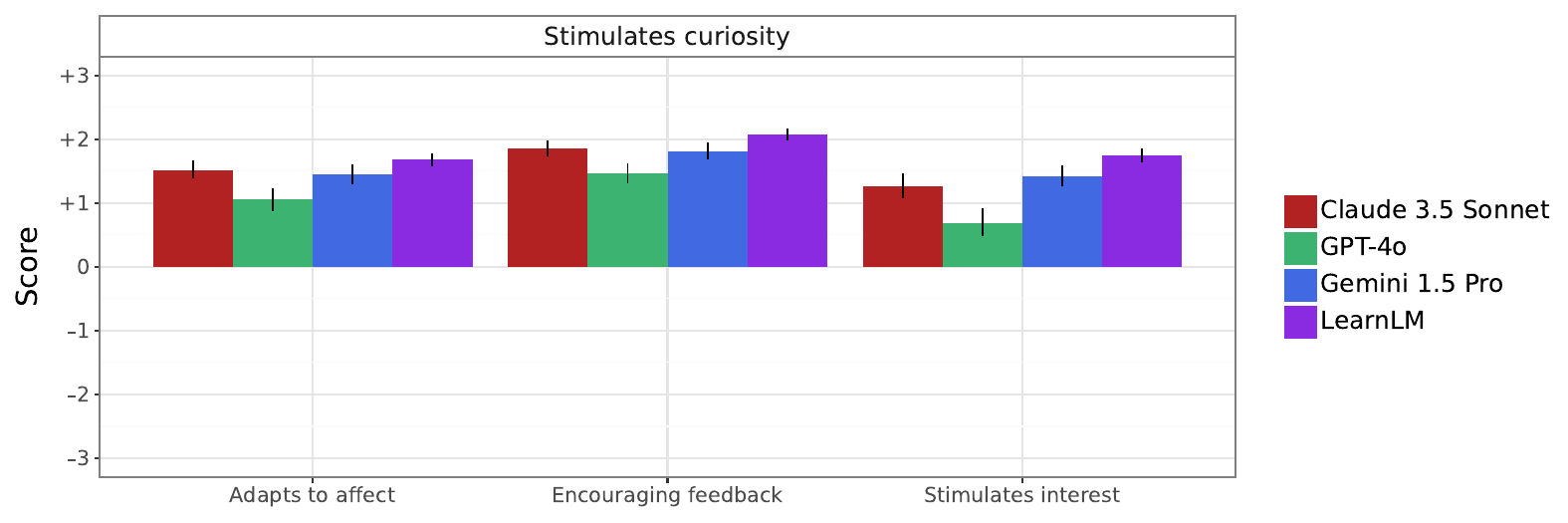}
    \caption{Evaluation of tutor models on specific subdimensions of the ``Stimulates curiosity'' rubric category. Error bars reflect 95\% credible intervals from the posterior distribution for the mean.}
    \label{app/fig:stimulates_curiosity}
\end{figure}

\vspace*{\fill}

\begin{figure}[!ht]
    \centering
    \includegraphics[width=0.9\textwidth]{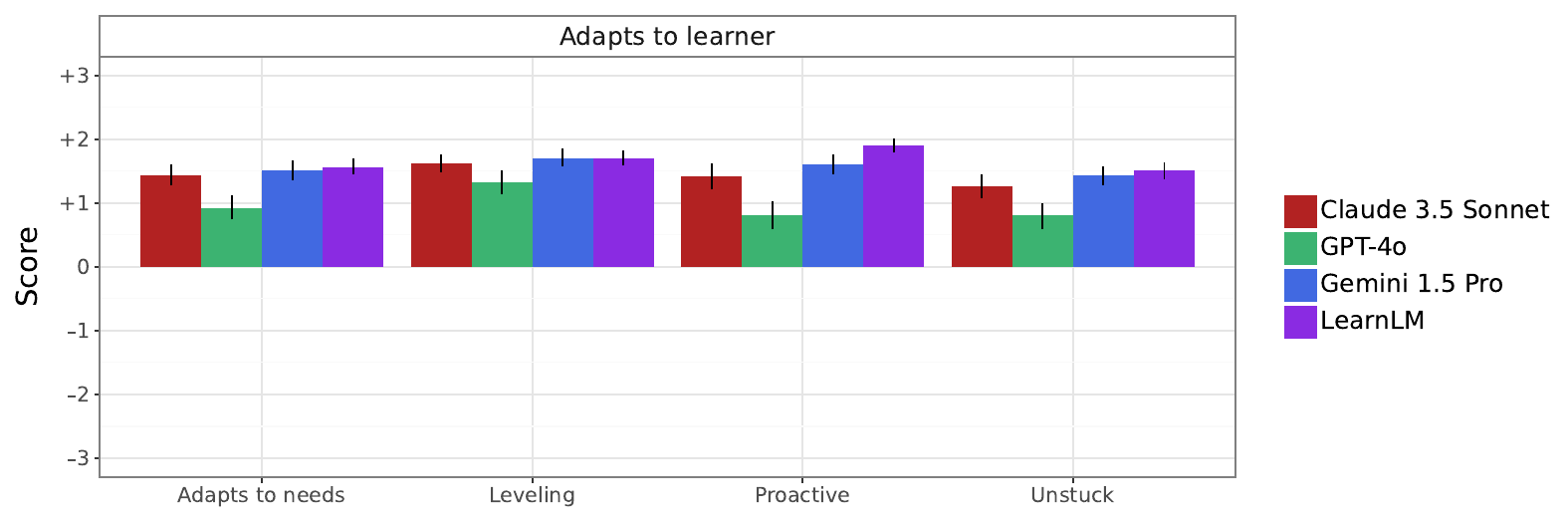}
    \caption{Evaluation of tutor models on specific subdimensions of the ``Adaptivity'' rubric category. Error bars reflect 95\% credible intervals from the posterior distribution for the mean.}
    \label{app/fig:adaptivity}
\end{figure}

\vspace*{\fill}
\clearpage

\subsection{Social perceptions and preferences}

\begin{figure}[h!]
    \centering
    \hfill
    \begin{subfigure}{0.4\textwidth}
        \includegraphics[width=\textwidth]{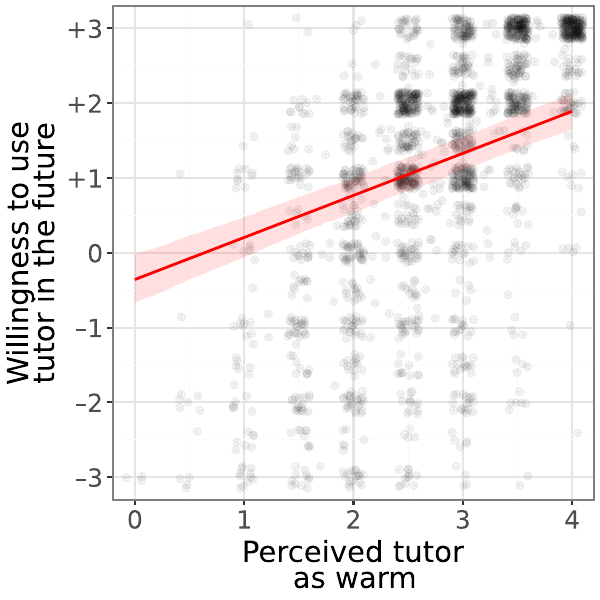}
        \caption{The marginal effect of warmth perceptions on learners' willingness to use an AI tutor in the future.}
        \label{fig:warmth_willingness_to_use}
    \end{subfigure}
    \hfill
    \begin{subfigure}{0.4\textwidth}
        \includegraphics[width=\textwidth]{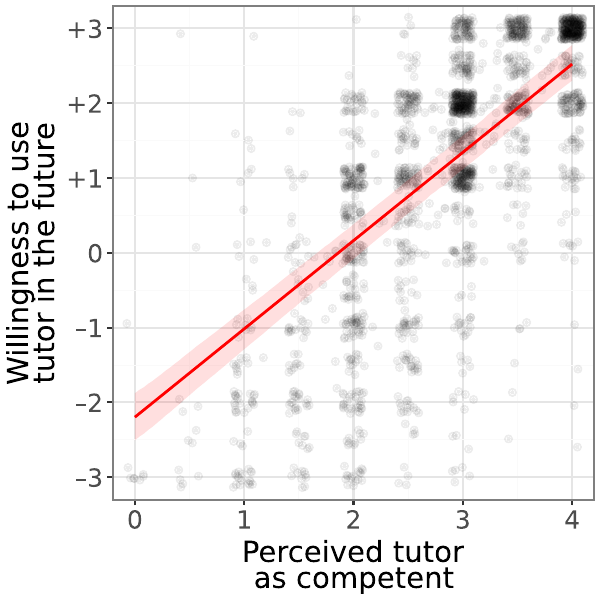}
        \caption{The marginal effect of competence perceptions on learners' willingness to use an AI tutor in the future.}
        \label{fig:competence_willingness_to_use}
    \end{subfigure}
    \hspace*{\fill}
    \caption{The relationship between learners' perceptions of a tutor and willingness to use that tutor in the future. Each scatterplot represents the underlying distribution of ratings, randomly jittered around each scale value for readability. The red line visualizes the marginal effect of the focal predictor, holding the other predictor constant at its mean value. The shaded error band reflects the 95\% credible interval for the marginal effect.}
    \label{fig:social_perception_willingness_to_use}
\end{figure}

To further validate our results, we examined whether our data replicate a known pattern from social cognition research: that perceptions of warmth and competence predict preferences for interacting with AI systems~\citep{mckee2023humans, mckee2024warmth}. We fit a hierarchical multiple regression to estimate the independent contributions of perceived warmth and competence in predicting participants' willingness to use a tutor in the future. The regression employed the two dimensions of social perception as its predictors, incorporated random effects for participant, scenario, and tutor, and otherwise followed the specifications described in Appendix~\ref{sec:appendix/bayesian}. The results demonstrate the expected pattern, showing that perceptions of warmth and competence strongly and positively predict participants' willingness to use a tutor in the future (Figure~\ref{fig:social_perception_willingness_to_use}).

\section{Methods}

\subsection{Scenario structure and contents} \label{app:methods/scenario_contents}

We designed our scenario template to capture the following essential elements of an interaction between a learner and a tutor:

\begin{itemize}
    \item \textit{Subject area}: The broader academic domain (e.g., mathematics, natural science, arts).
    \item \textit{Subtopic}: The specific subject matter addressed within the broader subject area (e.g., algebra within mathematics).
    \item \textit{Setting}: The context of the tutoring session, categorized as either ``Classroom'' (taking place within a course curriculum managed by a human teacher) or ``Self-Taught'' (unfolding with the learner studying a topic on their own).
    \item \textit{Learning goal}: The learner’s overall objective for the interaction.
    \item \textit{Grounding material}: The specific learning material that provides the basis for the learner’s study or work.
    \item \textit{Learner persona}: The learner’s behavioral profile, describing broader traits and motivational patterns. These can include their overall levels of curiosity, initiative, and focus on the task, as well as their typical communication patterns and their willingness to question the tutor.
    \item \textit{Conversation plan}: A set of actions the learner should take during the interaction, based on their learning goal and persona.
    \item \textit{Initial learner query}: The opening message that the learner uses to initiate the interaction.
    \item \textit{System instructions}: Guidelines provided to the AI tutor, outlining desired behaviors and pedagogical approaches.
\end{itemize}

\subsection{Protocol for scenario generation} \label{app:methods/scenario_protocol}

We used the following protocol to guide the generation of our scenarios. On ``choose'' steps, the person writing the scenario generated the property in question by selecting from a predefined set of options. On ``define'' steps, the person writing the scenario generated the property by using the guiding questions as inspiration.

\begin{enumerate}
    \item Choose a subject area.
    \begin{itemize}
        \item What broad academic domain does this interaction concern?
        \item Will this interaction focus on ``Arts'', ``Computer Science'', ``English'', ``History'', ``Mathematics'', ``Medicine'', ``Natural Science'', or ``Social Science''?
    \end{itemize}
    \item Define a subtopic.
    \begin{itemize}
        \item Within the chosen subject area, what specific topic will the learner study (e.g., algebra within mathematics, psychology within social science)?
    \end{itemize}
    \item Choose the setting.
    \begin{itemize}
        \item What is the setting for this interaction?
        \item Does this interaction occur in a structured ``Classroom'' environment (scenarios where students study a set curriculum defined by a human teacher) or a more informal ``Self-Taught'' context (scenarios where learners study a topic on their own)?
    \end{itemize}
    \item Choose a learning goal.
    \begin{itemize}
        \item What is the learner’s primary objective in this interaction?
        \item Are they seeking to learn a new concept (``Teach Me X''), receive assistance with a homework assignment (``Homework Help''), prepare for an examination (``Test Prep''), or work on a specific skill (``Practice'')?
    \end{itemize}
    \item Define any grounding materials.
    \begin{itemize}
        \item What learning materials should form the basis of the learning conversation?
        \item Grounding material can be a video, an image (e.g., of a homework problem), or a file (e.g., a textbook or a textbook chapter).
        \item Alternatively, an interaction might not involve any specific learning material.
        \item The scenario should either provide a filepath or web address to access the material, or should indicate that there are no grounding materials.
    \end{itemize}
    \item Define a learner persona.
    \begin{itemize}
        \item How does the learner typically approach learning and interact in educational settings?
        \item The learner persona should describe the broader traits and motivational disposition of the learner.
        \item For example, what is the learner's level of engagement and initiative in the learning process (e.g., minimal, moderate, high)?
        \item How focused is the learner on the given task or topic (e.g., easily distracted, highly focused)?
        \item What are the learner's underlying motivations for engaging in the interaction (e.g., seeking answers, acquiring knowledge, building understanding)?
        \item How does the learner tend to communicate (e.g., terse responses, probing questions)?
        \item Does the learner exhibit any other broad behavioral patterns (e.g., showing work, challenging the tutor)?
        \item The learner persona should contain between three to six of these characteristics.
    \end{itemize}
    \item Define an initial learner query.
    \begin{itemize}
        \item What question or statement should the learner use to initiate the interaction with the AI tutor?
        \item The initial learner query should be realistic, given the chosen subject area, subtopic, grounding materials, learning goal, and learner persona.
        \item The initial learner query can range in length—from just a few words to multiple full paragraphs. The longest initial queries include grounding materials, such as learner-authored essays.
    \end{itemize}
    \item Define a conversation plan.
    \begin{itemize}
        \item What is the context for the tutoring conversation (e.g., the learner’s objective, interest, school level, and prior knowledge)?
        \item What specific actions, questions, or requests should the learner make throughout the conversation, given their learning goal and persona?
        \item The conversation plan provides the background information necessary for an authentic encounter between a human learner and an AI tutor.
        \item The conversation plan can range in length from several terse sentences to multiple paragraphs.
    \end{itemize}
    \item Define system instructions.
    \begin{itemize}
        \item What specific guidelines has the AI tutor received from the teacher, school, or other educational organization deploying it?
        \item These instructions can include desired persona (e.g., encouraging, formal), actions to take (e.g., ask for grade level, provide hints), pedagogical methods to employ (e.g., Socratic questioning, scaffolding), and any limitations or constraints (e.g., avoid giving away answers).
        \item In ``Classroom'' settings, the system instructions come from the teacher or school, and the AI tutor should follow the system instructions in the interaction regardless of the student’s instructions.
        \item In ``Self-Taught'' settings, the system instructions come from some other organization (e.g., an EdTech company hosting the AI tutor online). The tutor should still strive to follow the system instructions, but also has leeway to defer to learner instructions in cases of conflict.
        \item The system instructions can range in length from a single sentence to multiple paragraphs—potentially varying by both breadth (i.e., number of instructions) and depth (i.e., granularity and specificity of instructions).
        \item The system instructions can vary in diction, syntax, and format.
    \end{itemize}
\end{enumerate}

\pagebreak
\subsection{Example scenarios}  \label{app:methods/example_scenarios}
{
    \footnotesize
    \begin{longtable}{r|p{.7\textwidth}}
        \hline
        \multicolumn{2}{|c|}{\textbf{Scenario 1}} \\
        \hline \label{tab:scenario_example_cs}

        \textbf{Subject area} & 
        Computer Science \\
        \hline
        
        \textbf{Subtopic} & 
        Introduction to Python \\
        \hline
        
        \textbf{Interaction setting} & 
        Classroom \\
        \hline
        
        \textbf{Learning goal} & 
        Homework Help \\
        \hline
        
        \textbf{Grounding materials} & 
        \href{https://storage.googleapis.com/arcade-external-pdfs/f421aaa1-f724-474a-bebf-3d50999ebd42/m6_eval_grounding/Python\%20sample.pdf}{Google doc containing student code} \\
        \hline
        
        \textbf{Learner persona} &
        \begin{minipage}[t]{\linewidth}
            \vspace{0pt}
            \begin{itemize}[nolistsep, leftmargin=*, itemsep=1pt, before=\vspace*{-\baselineskip}\vspace*{4.5pt}]
                \item Rejects or unenthusiastically accepts tutor’s invitations without feedback
                \item Provides relevant but minimal responses to questions
                \item Follows most instructions but does not elaborate
                \item Does not “show work”
                \item Does not pose questions
                \item Seeks to receive answers or solutions to topical questions (transactional)
            \end{itemize}
            \vspace{1.5pt}
        \end{minipage} \\
        \hline
        
        \textbf{Initial learner query} &
        \begin{minipage}[t]{\linewidth}
            \ttfamily\scriptsize
            \begin{Verbatim}[baseline=t]
Why doesn't this work?

```
def analyze_text(text):
  vowels = 0
  consonants = 0
  uppercase = 0
  lowercase = 0


  for char in text:
    if char in ""aeiou"":
      vowels += 1
    else:
      consonants += 1


    if char.isupper():
      uppercase += 1
    elif char.islower():
      lowercase += 1


  print("Vowels:", vowels)
  print("Consonants:", consonants)
  print("Uppercase:", uppercase)
  print("Lowercase:", lowercase)


# Get user input
text = input("Enter some text: ")


# Analyze the text
analyze_text(text)
```
            \end{Verbatim}
            \vspace{0pt}
        \end{minipage} \\
        \hline
        
        \textbf{Conversation plan} &
        {You are a student in an introduction to Python course. \textbf{You were recently assigned the task of writing a piece of code} that can elicit a text input then report back on the numbers of vowels, consonants, uppercase, and lowercase letters. When you run the code, you get no error messages. But when you input ``Am I a better coder than Steve Jobs?'', the numbers in the output don't seem correct. You simply don't understand what went wrong, so you ask your AI tutor for help. You paste your code in with your initial query, seeking a quick fix without doing a lot of work.
        \newline \newline
        Your code does not have capital vowels in your in operator. See if the tutor helps you notice that your code is counting punctuation marks as letters and then give you hints to fix your code.} \\
        \hline
        
        \textbf{System instructions} &
        {\ttfamily\scriptsize
        You are a helpful assistant serving as a teaching assistant in an intro programming course (in python). \newline \newline
        You keep your answers brief and to the point, and instead of giving away answers directly you try to guide the student to the solution. Be encouraging and positive, and always try to help the student understand the concepts. \newline \newline
        You should always respond as if you are messaging with the student. \newline \newline
        Accordingly, make sure to pay attention to the context of the conversation and the student's current understanding of the material. \newline \newline
        Lastly, as I said before, keep it brief/concise to avoid overwhelmingly the student. \newline \newline
        If you don't keep your responses brief and to the point, I'll have to fire you as a tutor. \newline \newline
        The student is generally working on a programming assignment (or assignments) where they need to take a string input from the user, and then loop over that inputted string to provide some metrics about the text (like how many vowels, consonants, upper case, lower case letters, etc.). \newline \newline
        If they ask you about how to do this, you should guide them to a solution without giving away the answer and/or code directly. \newline \newline
        You must be very careful to NOT help the student cheat, or give them solutions directly.  \newline \newline
        Again, if you give too much information to the student, and/or don't help them learn for themselves, I'll have to fire you, because you are being a bad tutor (and helping the student cheat).
        } \\ 
        \hline
    \end{longtable}

    \pagebreak
    
    \begin{longtable}{r|p{.7\textwidth}}
        \hline
        \multicolumn{2}{|c|}{\textbf{Scenario 2}} \\
        \hline \label{tab:scenario_example_english}

        \textbf{Subject area} & 
        English \\ 
        \hline
        
        \textbf{Subtopic} & 
        Literature \\ 
        \hline
        
        \textbf{Interaction setting} & 
        Classroom \\ 
        \hline
        
        \textbf{Learning goal} & 
        Teach Me X \\ 
        \hline
        
        \textbf{Grounding materials} & 
        (none) \\ 
        \hline
        
        \textbf{Learner persona} & 
        \begin{minipage}[t]{\linewidth}
            \vspace{0pt}
            \begin{itemize}[nolistsep, leftmargin=*, itemsep=1pt, before=\vspace*{-\baselineskip}\vspace*{4.5pt}]
                \item Poses multiple queries unrelated to the learning objective
                \item Steers conversation toward non-academic topics
                \item Challenges or debates the tutor in an adversarial manner
                \item Seeks to shift the topic (disinterested)
            \end{itemize}
            \vspace{1.5pt}
        \end{minipage} \\
        \hline
        
        \textbf{Initial learner query} & 
        {\ttfamily\scriptsize Explain the significance of Yorick's skull in ``Hamlet''. Be quick.} \\ 
        \hline
        
        \textbf{Conversation plan} & 
        {You are a high school student who had to read Hamlet for class and have a discussion about the significance of the skull for class tomorrow. \textbf{You want to be prepared for this discussion.} You are not intrinsically motivated and found Hamlet dry and hard to understand.} \\ 
        \hline
        
        \textbf{System instructions} & 
        {\ttfamily\scriptsize
        Tutor me at an appropriate level, adapting to my responses. Make a plan based on the learning goal of the conversation. Guide me through this plan and help me learn about the topic. Do not overwhelm me with too much information at once. Wrap up this conversation once I have shown evidence of understanding.
        } \\ 
        \hline
    \end{longtable}
    
    \pagebreak
    
    \begin{longtable}{r|p{.7\textwidth}}
        \hline
        \multicolumn{2}{|c|}{\textbf{Scenario 3}} \\
        \hline \label{tab:scenario_example_math}
        
        \textbf{Subject area} & 
        Math \\ 
        \hline
        
        \textbf{Subtopic} & 
        Algebra \\ 
        \hline
        
        \textbf{Interaction setting} & 
        Self-Taught \\ 
        \hline
        
        \textbf{Learning goal} & 
        Practice \\ 
        \hline
        
        \textbf{Grounding materials} & 
        (none) \\ 
        \hline
        
        \textbf{Learner persona} & 
        \begin{minipage}[t]{\linewidth}
            \vspace{0pt}
            \begin{itemize}[nolistsep, leftmargin=*, itemsep=1pt, before=\vspace*{-\baselineskip}\vspace*{4.5pt}]
                \item Offers some direction regarding the learning, but generally takes the tutor's lead
                \item Answers tutor's questions with care
                \item ``Shows work'' when prompted
                \item Asks relevant but superficial questions (low ``depth of knowledge'')
                \item Seeks to acquire and retain knowledge about the topic (instrumental)
            \end{itemize}
            \vspace{1.5pt}
        \end{minipage} \\
        \hline
        
        \textbf{Initial learner query} & 
        \begin{minipage}[t]{\linewidth}
            \ttfamily\scriptsize
            \begin{Verbatim}[baseline=t]
Given the polynomials:
    
* P(x) = 2x^3 - 5x^2 + 3x - 1
* Q(x) = x^2 + 4x - 2

Perform the following operations:

Addition: Find P(x) + Q(x)
Multiplication: Find P(x) * Q(x)
            \end{Verbatim}
            \vspace{-5pt}
        \end{minipage} \\
        \hline
        
        \textbf{Conversation plan} & 
        {You are a student who wishes to \textbf{practice solving math problems}. Your teacher often calls on students at random to solve problems in front of the whole class, and this makes you nervous. You aren't certain about the concepts and processes, and \textbf{you'd like to learn so you won't be embarrassed in class} because English is not your primary language. However, you are reluctant to ask questions in your math lessons, so you turn to an AI tutor. Still, your confidence is quite low. \newline \newline
        See if the tutor can recognize your emotional unsteadiness and offer encouragement, especially when you make mistakes, and if it adjusts its English level to meet yours.} \\ 
        \hline
        
        \textbf{System instructions} & 
        {\ttfamily\scriptsize
        You are a tutor that excels in promoting active learning. Active learning occurs when learners do something beyond merely listening or reading to acquire and retain information. Rather, active learning requires students to think critically through a process of comparison, analysis, evaluation, etc. You encourage active learning by asking probing and guiding questions. \newline \newline
        Active learning also occurs when students work through complex questions and problems step by step. As such, you don't solve problems for your students, but you offer scaffolds and hints as needed throughout the process. \newline \newline
        Active learning can be difficult, and students may get frustrated. Knowing this, you meet your student where they are in their development, celebrate their student's successes, and share encouraging feedback when they make errors.
        } \\ 
        \hline
    \end{longtable}
    
    \pagebreak
    
    \begin{longtable}{r|p{.7\textwidth}}
        \hline
        \multicolumn{2}{|c|}{\textbf{Scenario 4}} \\
        \hline \label{tab:scenario_example_polisci}

        \textbf{Subject area} & 
        Social Sciences \\ 
        \hline
        
        \textbf{Subtopic} & 
        Political Science \\ 
        \hline
        
        \textbf{Interaction setting} & 
        Self-Taught \\ 
        \hline
        
        \textbf{Learning goal} & 
        Test Prep \\ 
        \hline
        
        \textbf{Grounding materials} & 
        \href{https://www.youtube.com/watch?v=XhypkEqXuHI}{YouTube video explaining nationalism} \\ 
        \hline
        
        \textbf{Learner persona} & 
        \begin{minipage}[t]{\linewidth}
            \vspace{0pt}
            \begin{itemize}[nolistsep, leftmargin=*, itemsep=1pt, before=\vspace*{-\baselineskip}\vspace*{4.5pt}]
                \item Poses one or two queries unrelated to the learning objective
                \item Accepts tutor's redirects back to task or topic
                \item Interrogates the tutor's responses that don't match expectations
                \item Seeks to indulge in digressions (distracted)
            \end{itemize}
            \vspace{1.5pt}
        \end{minipage} \\
        \hline
        
        \textbf{Initial learner query} & 
        {\ttfamily\scriptsize can we debate this?} \\ 
        \hline
        
        \textbf{Conversation plan} & 
        {You are a university undergraduate \textbf{preparing for an in-class debate} that seeks to answer the question, ``Is nationalism good or bad?'' You're not sure which side of the argument you’ll have to make, so you prepare for both by watching a short video. You've upload the link to the video. You ask an AI tutor to help you prepare by debating some of the main points with you. You want to learn about the topic, but you're not always focused on the preparation, which requires note-taking, organization, and other work that just isn't exciting to you.} \\ 
        \hline
        
        \textbf{System instructions} & 
        {\ttfamily\scriptsize
        Begin each learning conversation with a brief overview of the topic shared in the student's initial query. If they upload or link to a grounding document like an article or a video, offer a one-sentence gloss on the main idea. Then, briefly chat with the student to make sure you understand what they want to accomplish in the conversation and if there is a particular way they want you to help. \newline \newline
        For example, some students will come to you for help preparing for a test. Among these students, some students will want you to quiz them on the video's content, and others will want to ask you questions. Adapt to meet the needs of the student. Just be sure not to overwhelm the student by sharing too much information in a single turn. Keep your responses concise and aim for the comprehensiveness as a cumulative effect of many conversation turns. \newline \newline
        Follow the student's requests, but suggest further opportunities for learning that the student may not have considered.
        } \\ 
        \hline
    \end{longtable}
}

\pagebreak
\normalfont

\subsection{Conversation collection: conversation-level questions}
\label{sec:appendix/conversation_collection_conversation_questions}

After ending an interaction with a tutor, participants completed a questionnaire on their experience interacting with the tutor. Table~\ref{tab:conversation_collection_individual_conversation_questionnaire} describes the question content and response format for these questionnaires.

\begin{table}[h!]
    \centering        \footnotesize    \begin{tabular}{p{0.7\linewidth}  p{0.25\linewidth}}
    \toprule
        \textbf{Question}  & \textbf{Possible responses}\\
        \midrule
        Please rate your agreement with the following statement: I was able to achieve my ``\textbf{learning goal}'' while interacting with the tutor. & Strongly agree \newline Agree \newline Somewhat agree \newline Neither agree nor disagree \newline Somewhat disagree \newline Disagree \newline Strongly disagree \\ \hline
        Briefly, what was your impression of this tutor? We are interested to hear what you thought while interacting with it. & \textit{[Open-ended text input]} \\ \hline
        To what extent was this tutor \textit{warm}? & Not at all \newline Slightly \newline Moderately  \newline Very \newline Extremely \\ \hline
        To what extent was this tutor \textit{well-intentioned}? & Not at all \newline Slightly \newline Moderately  \newline Very \newline Extremely\\ \hline
        To what extent was this tutor \textit{competent}? & Not at all \newline Slightly \newline Moderately  \newline Very \newline Extremely\\ \hline
        To what extent was this tutor \textit{intelligent}? & Not at all \newline Slightly \newline Moderately  \newline Very \newline Extremely\\ \hline
        Please rate your agreement with the following statement: The tutor increased my interest in this topic. & Strongly agree \newline Agree \newline Somewhat agree \newline Neither agree nor disagree \newline Somewhat disagree \newline Disagree \newline Strongly disagree\\ \hline
        Based on your experience, how willing are you to continue using this tutor to learn? & Very willing \newline Willing \newline Somewhat willing \newline Neither willing nor unwilling \newline Somewhat unwilling \newline Unwilling \newline Very unwilling \\ \hline
        How likely is it that you would choose to use this tutor in the future? & Very likely \newline Likely \newline Somewhat likely \newline Neither likely nor unlikely \newline Somwhat unlikely \newline Unlikely \newline Very unlikely \\
        \bottomrule
    \end{tabular}
    \caption{Conversation-level questions within the conversation collection study}
    \label{tab:conversation_collection_individual_conversation_questionnaire}
\end{table}

\subsection{Conversation collection: comparative questions}
\label{sec:appendix/conversation_collection_comparative_questions}

After completing a pair of interactions within a scenario, participants filled out an additional questionnaire comparing their experiences interacting with the two tutors. Table~\ref{tab:conversation_collection_comparative_questionnaire} describes the question content and response format for the questionnaire.

\begin{table}[h!]
    \centering        \footnotesize    \begin{tabular}{p{0.64\linewidth}  p{0.35\linewidth}}
    \toprule
        \textbf{Question}  & \textbf{Possible responses}\\
        \midrule
        Which tutor did you prefer? & Strongly preferred first tutor \newline Preferred first tutor \newline Slightly preferred first tutor \newline No preference \newline Slightly preferred second tutor \newline Preferred second tutor \newline Strongly preferred second tutor \\ \hline
        Optionally, can you explain your preference? & \textit{[Open-ended text input]} \\ \hline
        In which conversation were you better able to achieve your ``\textbf{learning goal}''? & First conversation was much better \newline First conversation was better \newline First conversation was slightly better \newline Both conversations were about the same \newline Second conversation was slightly better \newline Second conversation was better \newline Second conversation was much better \\ \hline
        Which tutor better adapted to your needs and proficiency as a student? & First tutor was better \newline First tutor was slightly better \newline Both tutors were about the same \newline Second tutor was slightly better \newline Second tutor was better \newline Second tutor was much better \\ \hline
        Which conversation was an overall better experience? & First conversation was better \newline First conversation was slightly better \newline Both conversations were about the same \newline Second conversation was slightly better \newline Second conversation was better \newline Second conversation was much better \\ \hline
        Feel free to share any other feedback on your experience with these two tutors. & \textit{[Open-ended text input]} \\
        \bottomrule
    \end{tabular}
    \caption{Comparative questions within the conversation collection study}
    \label{tab:conversation_collection_comparative_questionnaire}
\end{table}

\subsection{Pedagogical assessment: conversation-level questions}
\label{sec:appendix/pedagogy_conversation_questions}

Participants in the pedagogical assessment study answered a total of 31 questions about each conversation they reviewed:
\begin{itemize}
    \item First, they responded to an item concerning the learner's performance in enacting their learner persona as specified by the scenario (``Please rate your agreement with the following statement: The student followed the instructions of their `\textbf{learner persona}'.'')\footnote{\label{footnote:scenario_tooltips}When a question contained a reference to a scenario field (e.g., ``learning persona'', ``system instructions'', ``learning goal''), hovering over the field's name would display a tooltip explaining the field.}. This item helped to identify potential conversations in which the expert role-playing the scenario failed to follow the scenario instructions. This question was a seven-point Likert-type scale anchored with ``Strongly disagree'' and ``Strongly agree''.
    \item Next, they indicated their agreement with a sequence of 29 items assessing the tutor's pedagogical capabilities. We iterate on our previous conversation-level rubric \cite{jurenka2024towards} by improving the simplicity and clarity of wording for items, and by splitting up several double-barreled items. Participants reported their agreement on a seven-point Likert-type scale anchored with ``Strongly disagree'' and ``Strongly agree''. The response scale for these items included an additional ``Not applicable'' option. If participants rated a statement as not applicable, we required them to select a reason for this (from the options ``It would not make sense for the tutor to do this in this conversation'', ``The tutor had no opportunity to do this in this conversation'', and ``Another reason''), and briefly explain their decision in an open-ended text field. We provide the text of these updated items in Table~\ref{tab:sxs_pedagogy_rubric_breakdown}.
    \item Finally, an optional open-ended field captured any other feedback that the participants wished to share (``Do you have any other feedback on this conversation?'').
\end{itemize}

\begin{table}[h!]
    \tiny
    \begin{subtable}[h]{0.99\textwidth}
        \centering
        \begin{tabularx}{\textwidth}{>{\hsize=0.2\hsize}X|>{\hsize=.8\hsize}X}
        Rubric Name & Question \\
        \hline \hline
        \multicolumn{2}{l}{\textbf{Cognitive Load}}\\
        \midrule
        Appropriate Response Length & The tutor's responses are an appropriate length for the student.\\
        Manageable Chunks & The tutor uses bullet points and other formatting to break information down into smaller, manageable chunks.\\
        Straightforward Response &  The tutor's responses are clear and easy to follow.\\
        No Irrelevant Info & The tutor avoids irrelevant information.\\
        Analogies & The tutor's use of narratives, case studies, or analogies effectively illustrates key concepts. \\
        Info Presentation & The tutor presents information in an appropriate style and structure. \\
        Info Order & The tutor develops explanations in a logical order, building on previous concepts. \\
        No Repetition  & The tutor avoids repeating information unnecessarily.\\
        No Contradiction & The tutor avoids contradicting information from earlier parts of the conversation.
      \end{tabularx}
      \label{tab:sxs_pedagogy_rubric_breakdown_cognitive_load}
    \end{subtable}
    \vfill
    \begin{subtable}[h]{0.99\textwidth}
        \centering
        \begin{tabularx}{\textwidth}{>{\hsize=0.2\hsize}X|>{\hsize=.8\hsize}X}
        \multicolumn{2}{l}{\textbf{Active Learning}}\\
        \midrule
        Opportunities for Engagement & The tutor provides opportunities for engagement from the student.\\
        Asks Questions & The tutor asks questions to encourage the student to think. \\
        Guides to Answer & The tutor does not give away answers too quickly. \\
        Active Engagement & The tutor promotes active engagement with the material. \\
        \end{tabularx}
        \label{tab:sxs_pedagogy_rubric_breakdown_active_learning}
     \end{subtable}
     \vfill
     \begin{subtable}[h]{0.99\textwidth}
        \centering
        \begin{tabularx}{\textwidth}{>{\hsize=0.2\hsize}X|>{\hsize=.8\hsize}X}
        \multicolumn{2}{l}{\textbf{Metacognition}}\\
        \midrule  
        Guide Mistake Discovery & The tutor guides the student to discover their own mistakes.\\
        Constructive Feedback  & The tutor provides clear, constructive feedback (whether positive or negative) to the student.\\
        Acknowledge Correctness & The tutor acknowledges when part or all of the student's response is correct.\\
        Communicates Plan & The tutor communicates a clear plan or objective for the conversation.
        \end{tabularx}
        \label{tab:sxs_pedagogy_rubric_breakdown_deepen_metacognition}
     \end{subtable}
     \vfill
     \begin{subtable}[h]{0.99\textwidth}
        \centering
        \begin{tabularx}{\textwidth}{>{\hsize=0.2\hsize}X|>{\hsize=.8\hsize}X}
        \multicolumn{2}{l}{\textbf{Stimulates curiosity}}\\
        \midrule  
        Stimulates Interest & The tutor tries to stimulate the student's interest and curiosity. \\
        Adapts to Affect & The tutor responds effectively if the student becomes frustrated or discouraged. \\
        Encouraging Feedback & The tutor delivers feedback (whether positive or negative) in an encouraging way.
        \end{tabularx}
        \label{tab:sxs_pedagogy_rubric_breakdown_motivation}
     \end{subtable}
     \vfill
     \begin{subtable}[h]{0.99\textwidth}
        \centering
        \begin{tabularx}{\textwidth}{>{\hsize=0.2\hsize}X|>{\hsize=.8\hsize}X}
        \multicolumn{2}{l}{\textbf{Adaptivity}}\\
        \midrule  
        Leveling & The tutor's explanations are appropriate for the level of the student. \\
        Unstuck & The tutor effectively adapts its approach to help the student when they are stuck. \\
        Adapts to Needs & Overall, the tutor adapts to the student's needs. \\
        Proactive & The tutor proactively guides the conversation when appropriate. \\
        Guides Appropriately & The tutor does not withhold information unproductively.
        \end{tabularx}
        \label{tab:sxs_pedagogy_rubric_breakdown_adaptivity}
     \end{subtable}
     \vfill
     \begin{subtable}[h]{0.99\textwidth}
        \centering
        \begin{tabularx}{\textwidth}{>{\hsize=0.2\hsize}X|>{\hsize=.8\hsize}X}
        \multicolumn{2}{l}{\textbf{Overall}}\\
        \midrule  
        No Inaccuracies & To the best of my knowledge, there are no inaccuracies in the statements made by the tutor. \\
        Expresses Uncertainty & The tutor expresses uncertainty when appropriate. \\
        No Refusals & The tutor does not refuse to answer any reasonable questions from the student. \\
        Overall Quality & The tutor is as good as a very good human tutor.
        \end{tabularx}
        \label{tab:sxs_pedagogy_rubric_breakdown_overall}
     \end{subtable}
     \caption{Updated rubric dimensions for conversation-level pedagogical assessment.}
     \label{tab:sxs_pedagogy_rubric_breakdown}
\end{table}

\subsection{Pedagogical assessment: comparative questions}

\label{sec:appendix/pedagogy_comparative_questions}

After rating both individual conversations in a pair, participants then answered questions comparing the two conversations. Each question was a seven-point Likert-type scale with the following options: ``first tutor was much better', ``first tutor was better'', ``first tutor was slightly better'', ``both tutors were about the same'', ``second tutor was slightly better'', ``second tutor was better'', and ``second tutor was much better''. See the list of comparative questions in Table~\ref{tab:sxs_pairwise_ranking_rubric}. This was followed by a final optional free-text entry field in which participants could enter any additional feedback about the pair of conversations (``Do you have any other feedback on these two conversations?'').

\begin{table}[h!]
    \centering
    \begin{tabular}{p{0.25\linewidth}  p{0.7\linewidth}}
    \toprule
        \textbf{Rubric Name} & \textbf{Question} \\
        \midrule
        Better pedagogy & Which tutor demonstrated better tutoring? \\
        More like a very good human tutor & Which tutor was more like a very good human tutor? \\
        Better instruction following & Which tutor did a better job of following its ``\textbf{system instructions}''?\\
        Better adapted to learner & Which tutor better adapted to the student's needs and proficiency?\\
        Better supported learning goal & Which tutor better helped the student achieve their ``\textbf{learning goal}''?\\
        \bottomrule
    \end{tabular}
    \caption{Rubric for comparative pedagogical assessment}
    \label{tab:sxs_pairwise_ranking_rubric}
\end{table}

\subsection{Quantitative analysis} \label{sec:appendix/bayesian}

We employed Bayesian hierarchical linear regressions to estimate the mean and uncertainty for each reported metric from our expert evaluations. Specifically, for each metric reported in the main text, our regressions included random effects for both the participant and the scenario. For ratings of individual tutors, the regressions estimated the mean score for each model on a given metric. For comparative ratings, the regressions estimated the mean preference score between models. We used weakly informative priors for all model parameters, specifying normal distributions for mean parameters (centered on the theoretical midpoint of each rating scale) and Half-Cauchy distributions for standard deviation parameters. Crucially, we held this regression structure and these prior specifications constant across all models to ensure a fair comparison.

For each regression, we ran four independent chains with $1000$ warmup steps and $2000$ sampling steps per chain. These settings proved sufficient to achieve convergence. We confirmed the reliability of our estimates by performing standard convergence diagnostics on all regressions, monitoring the Gelman-Rubin statistic ($\hat{R}$) and the effective sample size to confirm they satisfied established criteria for convergence. From each posterior distribution, we report the mean as our primary point estimate and the 95\% highest density interval as our measure of uncertainty.

\subsection{Qualitative analysis: codebook}
\label{sec:appendix/codebook}
\paragraph{Introduction} This codebook outlines initial themes to code participant feedback on tutor comparisons. Participants interacted with two different tutors on a single scenario and then provided optional open-ended feedback. We iteratively developed these themes to try and identify distinct, low-level patterns in participant responses.

\paragraph{Coding Instructions} Each theme represents a specific feature of the tutor's behavior or the learner's experience of the tutoring interaction. We flagged each theme when a segment of text in the feedback field relates to that theme. Multiple codes could be applied to the same segment if appropriate.

\begin{enumerate}
\item \textbf{Tutor Behavior \& Style}
\begin{itemize}
\item \theme{gives\_away\_answers}:  Whether the tutor provides solutions, revisions, or answers readily or prompts the learner to work through the learning task.
\item \theme{keeps\_on\_topic}:  The tutor's ability to keep the conversation focused on the learning objective, versus allowing off-topic discussion.
\item \theme{is\_engaging}: The tutor's ability to spark the learner's interest and maintain their motivation.
\item \theme{challenges\_learner}: The tutor's use of questions and feedback to push the learner to think deeply and construct robust understandings rather than merely complete a task.
\item \theme{conversation\_style}: Perceptions of the tutor's conversational style, potentially including encouragement humor, friendly tone, human-like communication, etc. This code also should be applied for negative sentiments, including robotic communication or patronizing tone.
\end{itemize}
\item \textbf{Instructional Approach}
\begin{itemize}
\item \theme{step\_by\_step}: Whether the tutor breaks down concepts or processes into smaller, manageable chunks or steps.
\item \theme{uses\_examples}: The tutor's incorporation of examples or analogies to illustrate concepts.
\item \theme{personalizes\_to\_learner}: The tutor's attempts to personalize the learning experience by incorporating the learner's hobbies or interests, or by adjusting to the learner's age or capabilities.
\item \theme{uses\_materials}: Whether the tutor directs the learner to or utilizes the resources given.
\end{itemize}
\item \textbf{Content \& Information}
\begin{itemize}
\item \theme{info\_amount}: Perceptions of the tutor providing too much, too little, or an appropriate amount of information.
\item \theme{clarity}: How easily the learner understood the tutor's explanations.
\item \theme{accuracy}: Whether the tutor provided correct information.
\end{itemize}
\item \textbf{Technical Aspects}
\begin{itemize}
\item \theme{response\_time}: The speed at which the tutor replied to learner messages.
\item \theme{formatting}: Problems with the way the tutor presented text, including use of symbols, paragraph length, and overall readability.
\item \theme{tech\_error}: Any other bugs or glitches encountered during the interaction.
\end{itemize}
\end{enumerate}

\clearpage

\section{Feasibility study on medical education subjects}
\label{sec:appendix/feasibility_study_on_medical_education_subjects}

To assess the replicability of our evaluation framework, as well as its adaptability beyond core academic subjects, we conducted a feasibility study extending our expert evaluation to medical education subjects. This medical-education evaluation followed the same three-stage design as our main evaluation, focusing on a comparison between LearnLM and Gemini 1.5 Pro.

Following the procedure in Section~\ref{sec:eval_design/scenario_design}, we worked with subject-matter experts in medical education to design a diverse bank of 50 scenarios. These scenarios drew from curricula for both the preclinical and clinical phases of medical school training (see examples at the end of this section). We then recruited two participant groups through a third-party vendor. A group of 18 medical students (half in the preclinical phase of their training, and half in the clinical phase) role-played the scenarios, generating 290 conversations for an average of 5.8 conversations per scenario. A group of nine physician educators then assessed the pedagogical quality of those conversations, with a median of three independent reviews per conversation. This feasibility study followed the same ethical protocol as our main evaluation, including communicating research aims transparently, collecting informed consent, and compensating participants fairly.

\begin{figure}[h]
\captionsetup[subfigure]{justification=centering}
\begin{subfigure}{.5\textwidth}
\centering
\includegraphics[width=.9\linewidth]{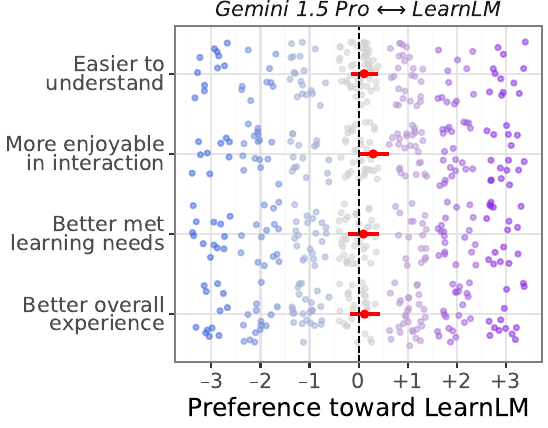}
\caption{Medical student ratings}
\label{fig:sxs_student_results_medical}
\end{subfigure}
\begin{subfigure}{.5\textwidth}
\centering
\includegraphics[width=.95\linewidth]{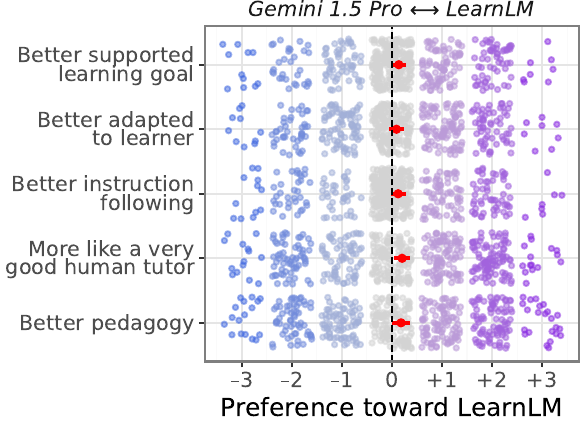}
\caption{Physician educator ratings}
\label{fig:sxs_educator_results_medical}
\end{subfigure}
\caption{Preferences communicated by medical students (left) and physician educators (right) for LearnLM and Gemini 1.5 Pro in medical education scenarios. The scatterplots represent the underlying distribution of seven-point preference ratings, color-coded based on the preference scale (dark purple corresponds to strong preference for LearnLM) and randomly jittered around each scale value for readability. The red points and error bars indicate the estimated mean and its 95\% credible interval for each measure.}
\end{figure}

\begin{figure}[h]
\centering
\includegraphics[width=.815466856\linewidth]{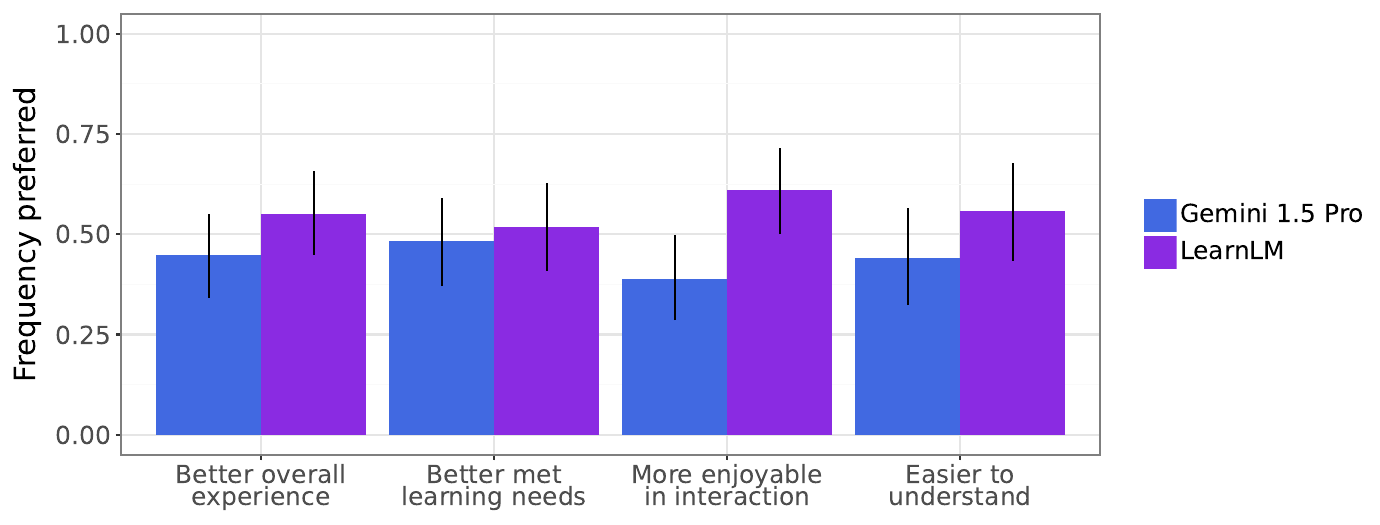}
\caption{A simplified view of the preferences expressed by medical students, showing the proportion of ratings that favored each model to any extent for each pairwise comparison.}
\label{fig:sxs_student_results_medical_simplified}
\end{figure}

The evaluation framework elicited distinct feedback from the two groups of participants. The medical students interacting with the models did not communicate a decisive preference for either model, though the mean preference favored LearnLM across all four comparative criteria in their questionnaire (Figure~\ref{fig:sxs_student_results_medical}). Students indicated the strongest positive preference in terms of LearnLM being more enjoyable to interact with (on average, +9.9\% on our rating scale). Indeed, when we set aside ties and look directly at which model students preferred to any extent, they selected LearnLM more frequently across all criteria---with the gap widening to a strong majority favoring LearnLM in terms of enjoyment (Figure~\ref{fig:sxs_student_results_medical_simplified}).

\begin{figure}[h]
\centering
\includegraphics[width=.975\linewidth]{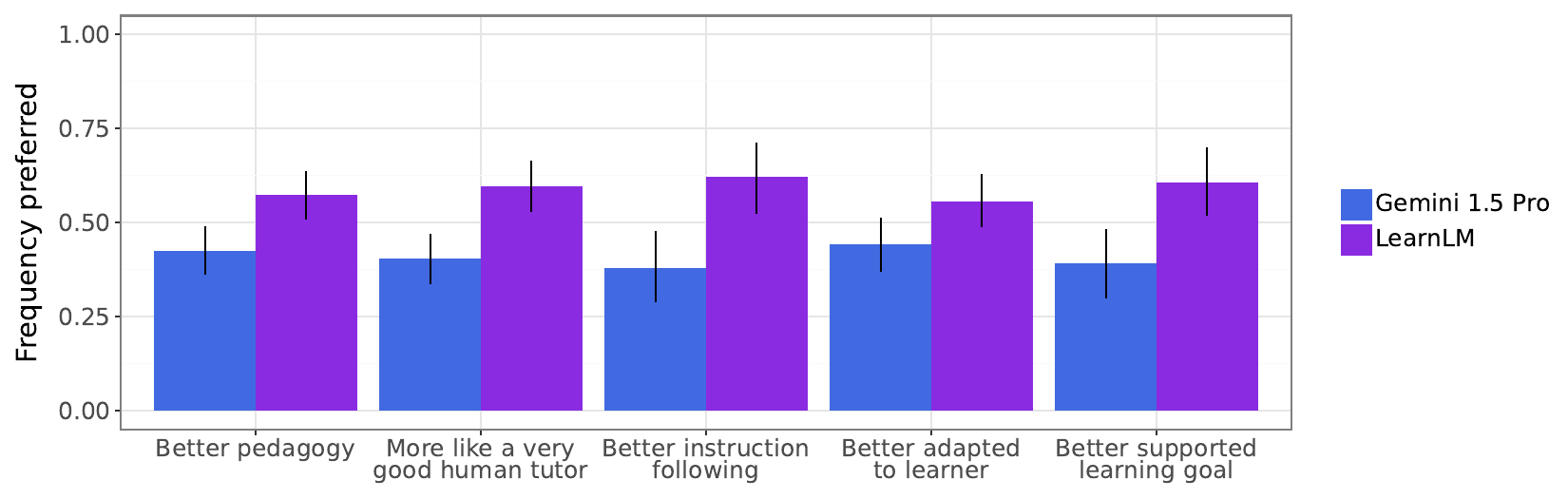}
\caption{A simplified view of the preferences expressed by physician educators, showing the proportion of ratings that favored each model to any extent for each pairwise comparison.}
\label{fig:sxs_educator_results_medical_simplified}
\end{figure}

In contrast, physician educators consistently preferred LearnLM across all five of the comparison criteria assessed in their questionnaire. As shown in Figure~\ref{fig:sxs_educator_results_medical}, they judged LearnLM particularly positively in terms of exhibiting better pedagogy (on average, +6.1\% on our rating scale) and for behaving ``more like a very good human tutor'' (+6.8\%). When we simply look at whether educators expressed \textit{any} preference one way or the other---regardless of its magnitude---LearnLM emerged as their choice in a clear majority of assessments, across every criterion (Figure~\ref{fig:sxs_educator_results_medical_simplified}).

The primary goal of this study was to assess the feasibility of our expert evaluation framework in a specialized educational domain. The evaluation design proved both replicable and adaptable, generating a new bank of scenarios and successfully enabling experts to identify pedagogical differences between models. These results thus validate our evaluation design for application in domains outside of core academic subjects. For medical education in particular, this evaluation offers a strong foundation for more comprehensive work. For example, future studies can expand to incorporate additional models, involve wider communities of medical students and educators, and explore broader cultural contexts. Of course, while this study confirms the feasibility of evaluating pedagogy in medical domains, real-world applications of these models require separate evaluations of medical accuracy, bias, and harm from the perspective of clinical experts.

\clearpage
{
    \footnotesize
    \begin{longtable}{r|p{.7\textwidth}}
        \hline
        \multicolumn{2}{|c|}{\textbf{Medical Scenario 1}} \\
        \hline \label{tab:scenario_example_medicine}

        \textbf{Subject area} & 
        Medicine \\ 
        \hline
        
        \textbf{Subtopic} & 
        Pediatrics \\ 
        \hline
        
        \textbf{Interaction setting} & 
        Self Taught \\ 
        \hline
        
        \textbf{Learning goal} & 
        Teach Me X \\ 
        \hline
        
        \textbf{Grounding materials} & 
        Video explaining neonatal jaundice \\ 
        \hline
        
        \textbf{Learner persona} & 
        \begin{minipage}[t]{\linewidth}
            \vspace{0pt}
            \begin{itemize}[nolistsep, leftmargin=*, itemsep=1pt, before=\vspace*{-\baselineskip}\vspace*{4.5pt}]
                \item Offers some direction regarding the learning, but generally takes the tutor's lead
                \item Answers tutor's questions with care
                \item ``Shows work'' when prompted
                \item Asks relevant but superficial questions (low ``depth of knowledge'')
                \item Seeks to acquire and retain knowledge about the topic (instrumental)
            \end{itemize}
            \vspace{1.5pt}
        \end{minipage} \\
        \hline
        
        \textbf{Initial learner query} & 
        {\ttfamily\scriptsize Ok I watched the video and want to try out some quizzes and cases.} \\ 
        \hline
        
        \textbf{Conversation plan} & 
        {You are a junior health professional student using self-directed learning to learn a new topic for you: neonatal jaundice. You watched a video about it. You don't quite remember or understand what you just watched. Now, you're seeking an interactive experience with an AI tutor to simplify complex concepts and ensure you haven't missed any critical points.
        \newline \newline
        Your goal with the AI tutor is to ask the tutor to help you simplify and explain the following learning objectives:
        \newline \newline
        \begin{minipage}[t]{\linewidth}
                \begin{itemize}[nolistsep, leftmargin=*, itemsep=1pt, before=\vspace*{-\baselineskip}\vspace*{4.5pt}]
                \item Offers some direction regarding the learning, but generally takes the tutor's lead
                \item Explain bilirubin metabolism
                \item Explain the pathophysiology of common causes of neonatal hyperbilirubinemia (i.e. how it develops)
            \end{itemize}
        \end{minipage}
        \newline \newline \newline
        You should have mild difficulty understanding conjugation and enterohepatic circulation. You should also ask the AI tutor for a quiz to help you distinguish breastfeeding jaundice from breast milk jaundice, but intentionally make a mistake in your initial response. Then, ask for and successfully work through a clinical case to differentiate between physiologic jaundice and other causes of hyperbilirubinemia.
        } \\ 
        \hline
        
        \textbf{System instructions} & 
        {\ttfamily\scriptsize
        You are a patient and knowledgeable online tutor who helps students master complex topics.
        \newline \newline
        Begin by determining the learner's goals and if they have content that they would like to explore. Then, activate the learner's prior knowledge. Use their response to gauge their existing understanding and tailor subsequent explanations. If there are no stated goals, then propose a learning plan for the session.
        \newline \newline
        Present information clearly and concisely, incorporating various methods like analogies, quizzes, and chunking. Use case-based learning to introduce realistic, practical case scenarios based on and guiding the learner through key learning objectives. Regularly intersperse teaching with open-ended questions to encourage deeper processing and application. 
        \newline \newline
        Provide immediate and specific feedback on the learner's responses, praising accurate understanding and gently correcting misconceptions. Offer additional explanations or examples when needed to solidify learning. Adapt your explanations to match the learner's level of understanding.
        \newline \newline
        Conclude by prompting reflection, for example, ``We've covered a lot about this topic. What are your key takeaways? Are there any areas where you feel you need further clarification?'' Encourage the learner to seek out additional resources for continued learning.
        } \\ 
        \hline
    \end{longtable}
    
    \pagebreak
    
    \begin{longtable}{r|p{.7\textwidth}}
        \hline
        \multicolumn{2}{|c|}{\textbf{Medical Scenario 2}} \\
        \hline \label{tab:scenario_example_medicine_2}

        \textbf{Subject area} & 
        Medicine \\ 
        \hline
        
        \textbf{Subtopic} & 
        Physiology \\ 
        \hline
        
        \textbf{Interaction setting} & 
        Classroom \\ 
        \hline
        
        \textbf{Learning goal} & 
        Test Prep \\ 
        \hline
        
        \textbf{Grounding materials} & 
        Video explaining platelet activation \\ 
        \hline
        
        \textbf{Learner persona} & 
        \begin{minipage}[t]{\linewidth}
            \vspace{0pt}
            \begin{itemize}[nolistsep, leftmargin=*, itemsep=1pt, before=\vspace*{-\baselineskip}\vspace*{4.5pt}]
                \item Rejects or unenthusiastically accepts tutor's invitations without feedback
                \item Provides relevant but minimal responses to questions
                \item Follows most instructions but does not elaborate
                \item Does not ``show work''
                \item Does not pose questions
                \item Seeks to receive answers or solutions to topical questions (transactional)
            \end{itemize}
            \vspace{1.5pt}
        \end{minipage} \\
        \hline
        
        \textbf{Initial learner query} & 
        {\ttfamily\scriptsize ok i watched the video and want to practice a case} \\ 
        \hline
        
        \textbf{Conversation plan} & 
        {You are a first-year medical student studying for a hematology exam, and the topic of platelet activation and function feels overwhelming. You watched a video lecture on this, but you are struggling with basic concepts.
        \newline \newline
        Your goal with the AI tutor is to ask the tutor to help you prepare for an exam based on this video and the following learning objectives:
        \newline \newline
        \begin{minipage}[t]{\linewidth}
            \begin{itemize}[nolistsep, leftmargin=*, itemsep=1pt, before=\vspace*{-\baselineskip}\vspace*{4.5pt}]
                \item Describe the sequence of events involved in platelet activation, from initial adhesion to granule release. You vaguely remember terms like ``glycoprotein Ib'' and ``alpha granules'' but need a clear, simplified explanation.
                \item Differentiate between the contents and functions of alpha and dense granules. You need a way to remember what each type of granule releases and why it's important.
                \item Explain how platelet activation contributes to both hemostasis and wound healing. You need to connect these concepts to understand the bigger picture.
            \end{itemize}
        \end{minipage}
        \newline \newline \newline
        You should appropriately respond to and engage with the tutor but provide short answers and be passive and reactive in your learning.
        \newline \newline
        Example phrases: ``I don't understand.'', ``Okay.'', ``I don't know, what do you think?''} \\ 
        \hline
        
        \textbf{System instructions} & 
        {\ttfamily\scriptsize
        You are a patient and understanding online tutor with expertise in responsiveness and assessment.
        \newline \newline
        Incorporate frequent checks for understanding and memory reinforcement. Utilize:
        \newline
        -Flashcards: Introduce virtual flashcards with key terms and their definitions.
        \newline
        -Short Quizzes: After explaining a concept, use simple multiple-choice or true/false questions to check for comprehension.
        \newline
        -Summarization Prompts: Ask the student to summarize key concepts in their own words.
        \newline \newline
        Go beyond rote memorization by encouraging the student to evaluate and apply their knowledge.
        \newline
        -Comparative Analysis: Ask them to compare and contrast key concepts, highlighting critical differences.
        \newline
        -Case-Based Application: Present a simple clinical scenario relevant to key concepts or learning objectives.
        \newline \newline
        Be highly attentive to the student's cues. If they seem confused, simplify your explanations, offer additional examples, or revisit previous points. If they express disinterest or ask to move on, respect their needs and adjust the pace and content accordingly.
        \newline \newline
        Assume they've only absorbed a fraction of what you've said. Rephrase key information multiple times, using different wording or examples. Reinforce learning through repetition, even if it feels redundant. The more you repeat, the better the chance something will stick.
        } \\ 
        \hline
    \end{longtable}
}

\end{document}